\documentstyle[epsf,times]{mn2e}
   
\newcommand{\simgt}{\lower.5ex\hbox{$\; \buildrel > \over \sim \;$}}
\newcommand{\simlt}{\lower.5ex\hbox{$\; \buildrel < \over \sim \;$}}
\newcommand{\citet}[1] {\cite{#1}}
\newcommand{\citep}[1] {(\cite{#1})}
 \newcommand{\bm}[1]{\mbox{\boldmath$#1$}}
 
 \newcommand{\skaco}[1]{\langle{#1}\rangle}

\newcommand{\colskip}{@{\hspace{0.3in}}}

\newcommand{\baredth}{\;\overline{\raise1.0pt\hbox{$'$}\hskip-6pt
\partial}\;}
\newcommand{\edth}{\;\raise1.0pt\hbox{$'$}\hskip-6pt\partial\;}
\begin{document}
\onecolumn \title[Non-Gaussian Errors in Weak Lensing Surveys]{
The Impact of Non-Gaussian Errors on Weak Lensing Surveys
}

 \author[M. Takada \& B. Jain]
{ Masahiro Takada$^{1}$\thanks{E-mail: masahiro.takada@ipmu.jp} 
and Bhuvnesh Jain$^2$\thanks{E-mail: bjain@physics.upenn.edu}  \\
$^1$Institute for the Physics of Mathematics of the Universe (IPMU), 
The University of Tokyo, 
Chiba 277-8582, Japan\\
$^2$ Department of Physics
and Astronomy, University of Pennsylvania, 
Philadelphia, PA 19104, USA
} 

\pagerange{\pageref{firstpage}--\pageref{lastpage}}

\maketitle

\label{firstpage}

\begin{abstract}
The weak lensing power spectrum carries cosmological information via 
its dependence on the growth of structure and on geometric factors. 
Since much of the cosmological information 
comes from scales affected by nonlinear clustering, 
measurements of the lensing power spectrum can be degraded by
non-Gaussian covariances. Recently there have been conflicting
studies about the level of this degradation. 
We use the halo model to estimate it and 
include new contributions related to the finite size of lensing
surveys, following Rimes and Hamilton's study of 3D simulations.
We find that non-Gaussian correlations between different
multipoles can degrade the cumulative signal-to-noise
for the power spectrum amplitude 
by up to a factor of 2 (or 5 for a worst-case model that exceeds
current N-body simulation predictions). 
However, using an eight-parameter Fisher analysis we find that the marginalized
errors on individual parameters are degraded by less than $10\%$ 
(or $20\%$ for the worst-case model). The smaller degradation in parameter 
accuracy is primarily because: 
individual parameters in a high-dimensional parameter space 
are degraded much less than the volume of the
full Fisher ellipsoid; lensing involves projections along the line of sight,
which reduce the non-Gaussian effect; some of the cosmological
information comes from geometric factors which are not degraded at
all. We contrast our findings with those of Lee \& Pen (2008) who 
suggested a much larger degradation in information content. 
Finally, our results give a useful guide for exploring survey design by giving
the cosmological information returns for varying survey area, depth and 
the level of some systematic errors. 
\end{abstract}

\section{Introduction}

Over the last decade a concordance model has emerged in cosmology in
which about two-thirds of the energy density of the universe today may
be in the form of dark energy.  This explains
the observation that we reside in an accelerating universe (Riess et
al. 1998; Perlmutter et al. 1999).  Despite its importance to the
formation and evolution of the universe there are no compelling theories
that explain the energy density or the properties of dark energy.

To address questions about the nature of dark energy a number of
ambitious wide-field optical and infrared imaging surveys have been
proposed. These range from space-based missions in the optical and
infrared, such as the Supernova Acceleration Probe
(SNAP\footnote{http://www.snap.lbl.gov}, proposed as the space-based
Joint Dark Energy Mission (JDEM)), and the Dark Energy UNiverse Explore
(DUNE\footnote{http://www.dune-mission.net/}), to ground-based surveys
such as the Panoramic Survey Telescope \& Rapid Response System
(Pan-STARRS\footnote{http://pan-starrs.ifa.hawaii.edu}), the Dark Energy
Survey (DES\footnote{ http://www.darkenergysurvey.org}), the Subaru Weak
Lensing Survey (Miyazaki et al. 2006), the Large Synoptic Sky Survey
(LSST\footnote{http://www.lsst.org}) and others. Each of these missions
approaches the study of dark energy using multiple, complementary
observational probes: gravitational weak lensing (WL) to study the
growth of structure and geometry, baryon oscillations to measure the
angular diameter distance vs. redshift relation, and Type Ia supernovae
to measure the luminosity distance vs. redshift relation. 

In this paper we focus on one of these probes, weak lensing or the
so-called cosmic shear, the bending of light by intervening mass
distribution that causes images of distant galaxies to be distorted
(e.g. Bartelmann \& Schneider 2001 for a thorough review).
These sheared source galaxies are mostly too weakly distorted for us to
measure the effect on single galaxies, but require large surveys containing 
millions of galaxies to detect the signal in a statistical way.  
The conventional method used for measuring cosmic shear
is the two-point correlation function whose Fourier-transform
is the shear power spectrum. Cosmic shear
correlations have been observed by various groups and used to constrain
cosmological parameters (most recently by Fu et al. 2008 
using the Canada-France-Hawaii Telescope Legacy Survey (CFHTLS)).

Lensing tomography refers to the use of depth information in the source
galaxies to get three-dimensional information about the lensing mass (Hu
1999; Huterer 2002; Heavens 2003; 
Takada \& Jain 2004; Song \& Knox 2004; Takada \& White 2004; 
also see Hoekstra \& Jain 2008 for a
recent review). By binning source galaxies in photometric redshift bins,
the evolution of the lensing power spectrum can be measured as a
function of redshift and angular scale.  This greatly improves the
sensitivity of lensing to the geometry of the universe as well as the
growth of mass clustering, both of which are sensitive to the nature of dark
energy.  This method has emerged as one of the most promising to obtain
precise constraints on the nature of dark energy if the systematic
errors are well under control (e.g., Albrecht et al. 2006).

Given the resources required for such surveys, it is important to
understand the statistical precision of cosmic shear observables and error
propagation in determination of cosmological parameters. 
Since cosmic shear probes the projected mass
distribution, the statistical properties of the cosmic shear field
reflect those of the mass distribution. For the case of the cosmic shear
power spectrum, its statistical precision is determined by the
covariance that contains three kinds of contributions: the shot noise
contamination due to intrinsic ellipticities, and the Gaussian and
non-Gaussian sample variances caused by the imperfect sampling of the
fluctuations (Scoccimarro et al. 1999; Cooray \& Hu 2001). The
non-Gaussian sample variance arises from the projection of the mass
trispectrum weighted with the lensing efficiency kernel. In fact most 
of the useful cosmological information contained in the lensing power
spectrum lies on small angular scales that are affected by nonlinear
clustering. Therefore non-Gaussian
errors can be significant in weak lensing measurements as indicated by a
few previous studies based on ray tracing simulations (White \& Hu
2000; Semboloni et al. 2007) and the halo model approach (Cooray \& Hu
2001; Pielorz 2008). However, the importance of non-Gaussian errors is not yet
fully understood, especially in terms of the prospects of future surveys
for constraining dark energy. This
will be also important in exploring optimal survey design for 
planned surveys.

Therefore, the aim of this paper is to study the covariances of the
lensing power spectrum based on the halo model approach (Cooray \& Sheth
2002 for a thorough review), and to estimate the impact of the
non-Gaussian errors on the power spectrum measurement as well as on the
determination of cosmological parameters.
We also study how the effect of the non-Gaussian errors
varies with survey parameters (depth and area) and in the presence of
systematic errors  -- photometric
redshift errors and shear calibration errors. Our analysis includes new
sources of non-Gaussian errors that inevitably arise for a finite
survey area, called the beat-coupling effect. 
This was pointed
out by Rimes \& Hamilton 2005 for the case of the 3D mass power spectrum
(also see Hamilton et al. 2006; Sefussati et al. 2006; Neyrinck
et al. 2006). More explicitly, if the scale of interest is embedded in a
large-scale (of order the survey size) overdensity or underdensity,
then 
the small scale fluctuations we want to measure may have 
grown more rapidly or
slowly than the ensemble average. This is predicted by
perturbation theory for gravitational clustering. This physical
correlations with the unseen large-scale fluctuations may add
uncertainties in measuring the power spectrum at scales of interest. 

Very recently, Lee \& Pen (2008) claimed that, by studying the angular
power spectrum of the SDSS galaxy distribution (i.e. not directly from 
lensing data), the effect
of non-Gaussian errors is very significant on angular scales of
$\sim 10$ arcminutes: they found that the cumulative signal-to-noise
ratio integrated over a range
of multipoles is two orders of magnitude lower than the case of the
Gaussian fluctuations.  Since galaxies are related to the lensing mass
fluctuations, does this imply that the non-Gaussian errors
significantly degrade the ability of lensing surveys to constrain cosmology?
This is the issue we would like to carefully address in this paper.

The structure of this paper is as follows. We define the lensing
power spectrum in the context of lensing tomography in
Section~\ref{sec:prelim} and 
the lensing covariances including the beat-coupling effect
in Section~\ref{sec:cov}. In Section~\ref{sec:results} we show the
results for the impact of non-Gaussian covariances on the power
spectrum measurement and parameter estimations. In
Section~\ref{sec:2ptcov} we study the covariances for the cosmic shear
correlation functions. Section~\ref{sec:discuss} is devoted to a 
discussion of our conclusions. 

\section{Preliminaries}
\label{sec:prelim}

\subsection{A CDM Model}

We will throughout this paper work in the context of a spatially flat
cold dark matter model for structure formation.  The expansion history
of the universe is given by the scale factor $a(t)$ in a homogeneous and
isotropic universe (e.g., see Dodelson 2003).  The expansion rate,
$H(t)\equiv \dot{a}(t)/a(t)$, is specified once the matter density
$\Omega_{\rm m0}$ (the cold dark matter plus the baryons) and dark
energy density $\Omega_{\rm de0}$ at present in units of the critical
density $3H_0^2/(8\pi G)$ are given, where $H_0=100~ h~{\rm km}~ {\rm
s}^{-1}~ {\rm Mpc}^{-1}$ is the Hubble parameter at present:
\begin{equation}
H^2(a)=H_0^2\left[\Omega_{\rm m0}a^{-3}+\Omega_{\rm de0}
e^{
-3\int^a_1 da' (1+w(a'))/a'}
\right],
\end{equation}
where we have employed the normalization $a(t_0)=1$ today and $w(a)$
specifies the equation of state for dark energy as $w(a)\equiv p_{\rm
de}(a)/\rho_{\rm de}(a)$.  Note that
$\Omega_{\rm m0}+\Omega_{\rm de0}=1$
and
$w=-1$ corresponds to a
cosmological constant.  The comoving distance $\chi(a)$ from an observer
at $a=1$ to a source at $a$ is expressed in terms of the Hubble
expansion rate as
\begin{equation}
\chi(a)=\int^1_a\!\!\frac{da'}{H(a')a^{\prime 2}}.
\end{equation}
This gives the distance-redshift relation $\chi(z)$ via the relation
$1+z=1/a$.

Next we need the redshift growth of density perturbations.  In linear
theory after matter-radiation equality, all Fourier modes of the mass
density perturbation, $\delta(\bm{x})(\equiv \delta
\rho_m(\bm{x})/\bar{\rho}_m)$, grow at the same rate, the growth rate
(e.g. see Eqn.~10 in Takada 2006 for details). Note that throughout this
paper we ignore effects of finite mass neutrinos and clustered dark
energy on the growth rate, causing a scale-dependent growth rate (e.g.,
Saito et al. 2008; Takada 2006).

\subsection{Tomographic Power Spectra of Cosmic Shear}

Gravitational shear can be simply related to the lensing convergence:
the weighted mass distribution integrated along the line of sight (e.g.,
see Mellier 1999; Bartelmann \& Schneider 2001; Schneider 2006 for
thorough reviews).  Photometric redshift information on source galaxies
allows us to subdivide galaxies into redshift bins, enabling more
cosmological information to be extracted, which is referred to as
lensing tomography (e.g., Hu 1999; Huterer 2002; Takada \& Jain 2004).
In the context of cosmological gravitational lensing the convergence
field with tomographic information is expressed as a weighted projection
of the three-dimensional mass density fluctuation field:
\begin{equation}
\kappa_{(i)}(\bm{\theta})=\int_0^{\chi_H}\!\!d\chi W_{(i)}(\chi)
\delta[\chi, \chi\bm{\theta}],
\label{eqn:kappa}
\end{equation}
where $\bm{\theta}$ is the angular position on the sky, $\chi$ is the
comoving angular diameter distance, and $\chi_H$ is the distance to the
Hubble horizon.  The lensing weight function $W_{(i)}(\chi)$ in the
$i$-th redshift bin, defined to lie between the comoving distances
$\chi_i$ and $\chi_{i+1}$, is given by
\begin{eqnarray}
W_{(i)}(\chi)&=&
\left\{
\begin{array}{ll}
{\displaystyle 
\frac{W_0}{\bar{n}_i}
a^{-1}\! (\chi)~ \chi
\int_{\chi_{i}}^{\chi_{i+1}}\!\!d\chi_{s}~ 
n_s(z)\frac{dz}{d\chi_s} \frac{\chi_{\rm s}-\chi}{\chi_s}},
& \chi\le\chi_{i+1},\\
0,& \chi>\chi_{i+1},
\end{array}
\right.
\label{eqn:weight}
\end{eqnarray}
where $W_0\equiv (3/2)\ \Omega_{\rm m0}H_0^2$ and $n_s(z)$ is the
redshift selection function of source galaxies, which is normalized as
$\int_0^{\infty}\!\!dz ~ n_s(z)= \bar{n}_{\rm g}$ with $\bar{n}_{\rm g}$
being the average number density of galaxies per unit steradian (see
around Eqn.~[\ref{eqn:ns}] in \S~\ref{sec:modelparas} for our definition
of the source redshift distribution).  Also $\bar{n}_i$ is the average
number density of sub-sample galaxies in the $i$-th redshift bin defined
as
\begin{equation}
\bar{n}_i=\int_{\chi_i}^{\chi_{i+1}}\!\!d\chi_s~ n_s(z)\frac{dz}{d\chi_s}. 
\label{eqn:ni}
\end{equation}

The cosmic shear fields are measurable only in a statistical sense.  The
most conventional methods used in the literature are the shear two-point
correlation function. The Fourier transformed counterpart is the shear
power spectrum.  For lensing tomography of $n_z$ redshift bins, using
the flat-sky and Limber's approximations (Limber 1954), there are
$n_z(n_z+1)/2$ spectra available:
\begin{equation}
P_{(ij)}(l)=\int_0^{\chi_H}\!\!d\chi W_{(i)}(\chi)
W_{(j)}(\chi)~ \chi^{-2}
P_\delta\!\left(k=\frac{l}{\chi}; \chi\right),
\label{eqn:ps}
\end{equation}
where $P_\delta(k)$ is the three-dimensional mass power spectrum. Note
that hereafter 
the quantities with subscript ``$\delta$'' denote those of the mass
density fluctuations.  For $l\simgt 100$ the major contribution to
$P_{(ij)}(l)$ comes from non-linear clustering (e.g., see Fig.~2 in
Takada \& Jain 2004).  We employ the fitting formula for the non-linear
$P_\delta(k)$ proposed in Smith et al.  (2003), assuming that it can be
applied to dark energy cosmologies by replacing the growth rate used in
the formula with that for a given dark energy model. We note in passing
that the issue of accurate power spectra for general dark energy
cosmologies still needs to be addressed carefully (Huterer \& Takada
2005; also see Ma 2007 for the related discussion).

As can be found form Eqn.~(\ref{eqn:ps}), the lensing power spectra
contain cosmological information via both the lensing efficiency kernel
and the mass clustering information contained in $P_\delta$, e.g. which
almost equally contribute to the final sensitivity to dark energy
parameters. On the other hand, the non-Gaussian errors of the lensing
fields arise from the non-Gaussianity of the nonlinear mass clustering
in structure formation. Thus, even if the underlying mass distribution
is highly non-Gaussian, the cosmological information the lensing carries
would be to some extent preserved via the lensing efficiency kernel,
which is one of notable differences from other methods such as the
galaxy power spectrum.

In reality, the observed power spectrum is contaminated by the intrinsic
ellipticity noise.  Assuming that the intrinsic ellipticity distribution
is uncorrelated between different galaxies, the observed power spectrum
between redshift bins $i$ and $j$ can be expressed as
\begin{equation}
P^{\rm obs}_{(ij)}(l)
=P_{(ij)}(l)+\delta^K_{ij}\frac{\sigma_\epsilon^2}{\bar{n}_{(i)}}
\label{eqn:psobs}
\end{equation}
where $\sigma_\epsilon$ is the rms intrinsic ellipticities per
component. 
Note that the Kronecker delta symbol $\delta^{K}_{ij}$ accounts
for the fact that the cross-spectra with $i\ne j$ are not contaminated
by the shot noise.  Therefore the shot noise contamination needs not be
subtracted from the estimated cross spectra that would reduce residual
uncertainties in practice.

\section{Covariances of the Cosmic Shear Power Spectra}
\label{sec:cov}

\subsection{Definition}
In reality the lensing power spectrum has to be estimated from the
Fourier or spherical harmonic coefficients of the observed lensing
fields constructed for a finite survey.  In this paper we assume the
flat-sky approximation and thus use Fourier wavenumbers $l$, which
are equivalent to spherical harmonic multipoles $l$ in the limit
$l \gg 1$ (Hu 2000).  Because the survey is finite, an infinite
number of Fourier modes are not available, and rather the discrete
Fourier decomposition has to be constructed in terms of the fundamental
mode that is limited by the survey size as will be in detail discussed
below.
We assume a homogeneous survey geometry for simplicity and do not
consider any complex boundary and/or masking effects. For this case, as
shown in Appendix B of Takada \& Bridle (2007), the lensing power
spectrum of the $a$-th multipole bin, $l_a$, may be estimated as
\begin{eqnarray}
P_{(ij)}^{\rm est}(l_a)&=&\frac{1}{\Omega_{\rm s}}
\int_{|\bm{l}'|\in l_a}
\!\!\frac{d^2\bm{l}'}{A(l_a)}
\tilde{\kappa}_{(i)\bm{l}'}\tilde{\kappa}_{(j)-\bm{l}'},
\label{eqn:psest_maintext}
\end{eqnarray}
where $\Omega_{\rm s}$ is the survey area, the integration range is
confined to the Fourier modes lying in the annulus of a given width,
$l_a-\Delta l_a/2\le l'\le l_a+\Delta l_a/2$ and $A(l_a)$ denotes the
integration area in the Fourier space approximately given by
$A(l_a)\equiv \int_{|\bm{l}'|\in l_a}d^2\bm{l}'\approx 2\pi l_a\Delta
l_a$ for the case of $l_a\gg \Delta l_a$. Throughout this paper we use
the subscripts $a,b$ to denote the multipole bins, while we use the
subscripts $i,j$ or $i',j'$ to denote the redshift bins.

Once an estimator of the lensing power spectrum is defined, it is
straightforward to compute the covariance (Scoccimarro et al 1999;
Cooray \& Hu 2001; Takada \& Bridle 2007). The covariance of cosmic
shear power spectra describes statistical uncertainties of the power
spectrum measurement for a given survey as well as how two spectra at
different multipole and/or redshift bins are correlated with each other.
Extending the formulation developed in Scoccimarro et al. (1999)
to the tomography case (Takada \& Bridle 2007), the
covariance matrix of the lensing power spectra is given by
\begin{eqnarray}
[\bm{C}]_{AB} &\equiv& \skaco{P^{\rm est}_{(ij)}(l_a)
P^{\rm est}_{(i'j')}(l_b)}-P_{(ij)}(l_a)P_{(i'j')}(l_b)
\equiv {\rm Cov}^{\rm Gaussian} + {\rm Cov}^{\rm NG}
\nonumber\\
&&\hspace{-2em}=\frac{\delta^K_{ab}}{(2l_a+1)\Delta lf_{\rm sky}}
\left[P^{\rm obs}_{(ii')}\!(l_a)P^{\rm obs}_{(jj')}\!(l_a)+
P^{\rm obs}_{(ij')}\!(l_a)P^{\rm obs}_{(ji')}\!(l_a)
\right]+\frac{1}{4\pi f_{\rm sky}}
\int_{|\bm{l}|\in l_a}\!\!\frac{d^2\bm{l}}{A(l_a)}
\int_{|\bm{l}'|\in l_b}\!\!\frac{d^2\bm{l}'}{A(l_b)}
T_{(iji'j')}(\bm{l},-\bm{l},\bm{l}',-\bm{l}'),
\label{eqn:pscov}
\end{eqnarray}
where $f_{\rm sky}$ is the sky coverage ($f_{\rm sky}=\Omega_{\rm
s}/4\pi$). We have shown two contributions to the covariance: 
the terms in the square brackets constitute 
the Gaussian contribution, and last term is the non-Gaussian term 
which is given by the lensing trispectrum $T$ defined as
\begin{equation}
\langle\kappa_{(i)}(\bm{l}_1)\kappa_{(j)}(\bm{l}_2)
\kappa_{(i')}(\bm{l}_3)\kappa_{(j')}(\bm{l}_4)\rangle
\equiv (2\pi)^2 \delta_D(\bm{l}_1+\bm{l}_2+\bm{l}_3+\bm{l}_4)
T_{(iji'j')}(\bm{l}_1,\bm{l}_2,\bm{l}_3,\bm{l}_4).
\end{equation}
In the Limber approximation, $T$ is a simple projection of 
the 3D mass trispectrum $T_\delta$ given as
\begin{eqnarray}
T_{(iji'j')}(\bm{l},\bm{l}^\prime,\bm{l}^{\prime\prime},
\bm{l}^{\prime\prime\prime})
=\int_0^{\chi_H}\!\!d\chi~
 W_{(i)}(\chi)W_{(j)}(\chi)W_{(i')}(\chi)W_{(j')}(\chi)
 \chi^{-6} T_{\delta}(\bm{k},\bm{k}^\prime,\bm{k}^{\prime\prime},
\bm{k}^{\prime\prime\prime}; \chi), 
\label{eqn:lenstrisp}
\end{eqnarray}
with $\bm{k}=\bm{l}/\chi$ and so on. Finally, the indices $A,B$ in the
covariance matrix of Eqn.~(\ref{eqn:pscov}) run over both multipole and
redshift bins. For 
tomography with $n_z$ redshift bins, there are
$n_z(n_z+1)/2$ different spectra available at each multipole. Hence, 
with $n_l$ multipole bins, the indices $A,B$ take values
$A,B=1,2,\dots,n_l\times n_z(n_z+1)/2$.
For example, for $n_z=3$ and $n_l=100$, 
the covariance matrix $\bm{C}$ has dimension $600\times600$.

The first term of the covariance matrix (second line on the r.h.s. of
Eqn.~[\ref{eqn:pscov}]) represents the Gaussian error contribution
ensuring that the two power spectra of different multipoles are
uncorrelated via $\delta^K_{ab}$, while the second term gives the
non-Gaussian errors includes correlation between power
spectra at different $l$'s.  
The two terms both scale with sky coverage as $\propto
1/f_{\rm sky}$ (but see below for an additional dependence in $T$). 
Note that the intrinsic ellipticity noise contributes
only to the Gaussian errors via $P_{(i)}^{\rm obs}$, as long as
intrinsic alignments of galaxy ellipticities are negligible.  
It should be
also noted that the non-Gaussian term does not depend on the multipole
bin width $\Delta l$ (because $\int\!\!d^2\bm{l}/A(l)\approx 1$), so
increasing $\Delta l$ only reduces the Gaussian contribution.
However, the signal-to-noise ratio and parameter
forecasts we will show below do not depend on the multipole bin width if
the bin width is not very coarse (also see Scoccimarro et al. 1999 for
the related discussion for the case of 3D mass power spectrum).

\begin{figure}
  \begin{center}
    \leavevmode\epsfxsize=13.cm \epsfbox{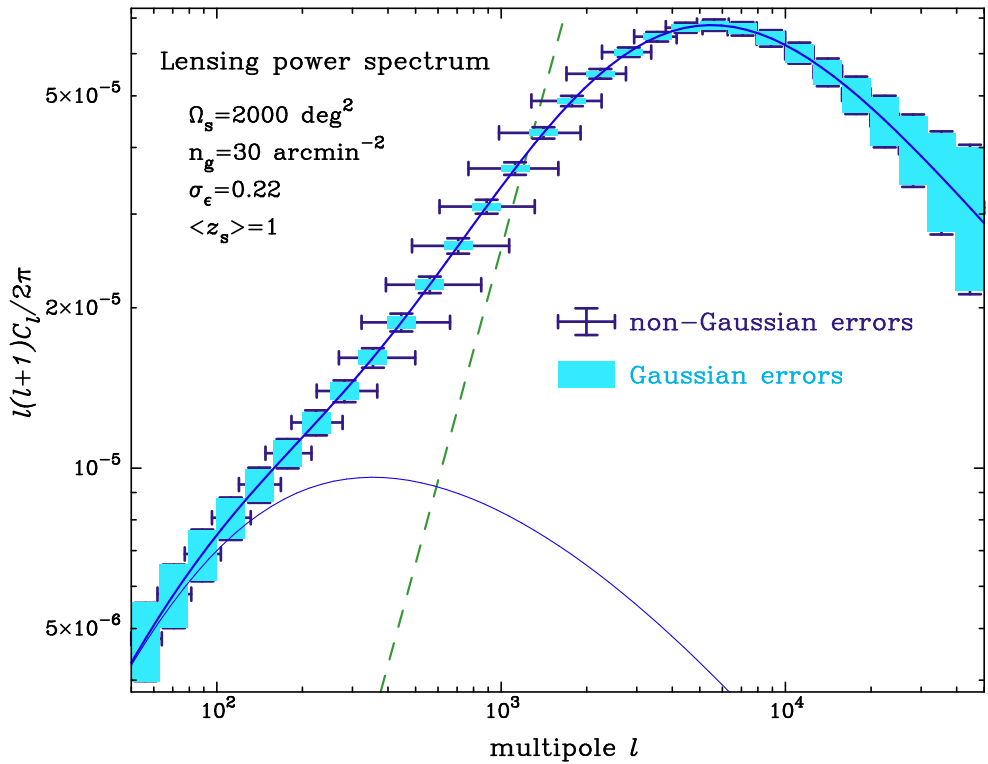}
  \end{center}
\caption{ The lensing power spectrum expected from our fiducial
2000 deg$^2$ ground-based survey.  
The bold solid curve shows the prediction for the
$\Lambda$CDM model (including non-linear evolution)
while the thin solid curve denotes the linear theory
prediction. The shaded boxes around the bold solid curve show the
expected measurement errors at each multipole bin
assuming Gaussian errors.  The vertical and horizontal error
bars around the bold solid curve show non-Gaussian effects. 
The horizontal error bars display correlations between neighboring
multipole bins caused by the non-Gaussian errors, while the vertical error
bars show the increase in errors compared to the Gaussian errors
(shaded boxes). 
} \label{fig:ps}
\end{figure}
Figure~\ref{fig:ps} gives a quick summary of the impact of
non-Gaussian errors on the shear power spectrum. 
The parameters for the cosmological model and lensing survey are
described in \S~\ref{sec:modelparas}. The shaded boxes around
the solid curve show the expected Gaussian errors on the power spectrum
for the assumed survey. The shot noise contribution to the Gaussian
errors becomes greater than the sample variance at wavenumbers $l$
greater than the intersection of the power spectrum points with the shot
noise line (dashed). The horizontal and vertical error bars
demonstrate the impact of the non-Gaussian errors on the power spectrum
measurements at each multipole bins. There are two effects. First, the
neighboring multipole bins are not independent: the width of each
horizontal error bar represents the range of multipoles where the
cross-correlation coefficient of power spectra (around the central
multipole in that bin) is greater than 0.1 (see 
\S~\ref{sec:pca} for the details).  Second, the vertical error
bars show the increase in the power spectrum measurement
uncertainties due to non-Gaussian errors. These are significant over the range
$100\simlt l\simlt 10^3$, where (a) nonlinear clustering is
important, and (b) sample variance dominates over shot
noise. 

\subsection{Effect of Finite Survey Area: 
Beat-Coupling Contribution}
\label{sec:bc}

As pointed out in Hamilton, Rimes \& Scoccimarro (2006; also see
Rimes \& Hamilton 2005 and Sefussati et al. 2006), there is
an additional contribution to the covariance arising from the imperfect
sampling of the Fourier modes due to a finite survey area, which we will
often refer as to the {\em beat-coupling} contamination. 
For a finite
survey of size $L$ (we will hereafter assume a survey area of
$\Omega_s=L^2$ for simplicity), the uncertainty principle tells us that we
cannot measure Fourier modes to a better accuracy than
$\varepsilon=2\pi/ L$: two modes that differ by $\bm{\varepsilon}$,
$\bm{l}\pm \bm{\varepsilon} $, cannot be distinguished due to the limited
resolution\footnote{Note that $\varepsilon$
for the fundamental mode is not to be confused with the intrinsic ellipticity,
$\epsilon$!}. All statistics we measure contains these
uncertainties.
Therefore we need to allow the trispectrum in the
covariance (\ref{eqn:pscov}) to have the uncertainties given as
\begin{eqnarray}
T(\bm{l}+\bm{\varepsilon},
-\bm{l}+\bm{\varepsilon}^\prime,\bm{l}'+\bm{\varepsilon}^{\prime\prime},
-\bm{l}'+\bm{\varepsilon}^{\prime\prime\prime})
\label{eqn:trispbc}
\end{eqnarray}
where all the wavevectors $\bm{\varepsilon}$ with/without primes denote the
fundamental modes with magnitude $|\bm{\varepsilon}|\simeq 2\pi/L$ but
can have different orientations.
Note that, since the power spectrum depends only on the {\em length} of
wavenumber $\bm{k}$ in a statistically homogeneous and isotropic
universe, an estimation of the power spectrum of wavenumber $l$ is not
affected by the uncertainty principle as long as it is smoothly
varying and $l \gg \varepsilon$:
\begin{equation}
P(|\bm{l}+\bm{\varepsilon}|)\simeq P(l). 
\end{equation}
Hence, the Gaussian error term in the covariance (\ref{eqn:pscov}) is
not affected by the beat-coupling contamination.  

For the non-Gaussian term, there is a non-vanishing
contribution due the beat-coupling effect. Even for $l \gg
\varepsilon$, nonlinear clustering
predicts non-vanishing correlations between the modes with wavevectors
$\bm{l}$ and $\bm{\varepsilon}$. Surprisingly, this effect yields additional
contribution to the covariance that are not negligible, as we show
below (and as shown by Rimes \& Hamilton 2005 for the 3D
power spectrum). 

\subsection{Halo Model Approach for the Covariance}
To compute the lensing power spectrum covariance using
Eqn.~(\ref{eqn:pscov}), we need to model the mass trispectrum. 
The model predictions need to describe the non-linear regime of
clustering because most of the useful 
cosmological information in lensing is contained on 
small angular scales. In this paper, we employ the dark matter halo
approach (Seljak 2000; Peacock \& Smith 2000; Ma \& Fry 2000;
Scoccimarro et al. 2001; also see Cooray \& Sheth 2002 for a review), where
the $n$-point correlations of the mass distribution are modeled in terms
of two separate contributions: correlation of dark matter particles
within one halo, and correlations of particles in different halos.  In
previous work we have found that, up to the 4-point correlation
functions, the halo model gives fairly accurate predictions to match
$N$-body simulation results at $\sim 10-30\%$ level (Takada \& Jain 2002,
2003a,b).  For our purpose, 
the halo model prescription is reasonably adequate; we discuss
possible improvements below.

Based on the halo model approach the mass trispectrum is given by
\begin{eqnarray}
T_\delta(\bm{k},\bm{k}^\prime,\bm{k}^{\prime\prime},\bm{k}^{\prime\prime\prime})
&=&
T_\delta^{\rm PT}
\!(\bm{k},\bm{k}^\prime,\bm{k}^{\prime\prime},\bm{k}^{\prime\prime\prime})
+T_\delta^{1h}
\!(\bm{k},\bm{k}^\prime,\bm{k}^{\prime\prime},\bm{k}^{\prime\prime\prime}),
\label{eqn:trisp_m}
\end{eqnarray}
where $T^{1h}_{\delta}$ denotes the 1-halo term and $T_\delta^{\rm PT}$
denotes the perturbation theory prediction (e.g., Makino et al. 1992;
Jain \& Bertschinger 1994; also see Bernardeau et al. 2002 for a
review). The details of our halo model implementation are given
in Takada \& Jain (2003a,b). We will calculate the two contributions
to $T_\delta$ given above, and thus estimate the non-Gaussian
covariance as the sum of two terms: ${\rm Cov}^{\rm NG} = {\rm
  Cov}^{\rm NG,PT} +  {\rm Cov}^{{\rm NG},1h}$. 

In Eqn.~(\ref{eqn:trisp_m}) we have dropped the 2- and 3-halo term
contributions. We have thus assumed 
that the full trispectrum is well approximated 
by the sum of the 1-halo term and the perturbation theory prediction. 
Our rationale for this approximation is: (1) The 1-halo term
is dominant in the highly non-linear regime,
while in the linear regime the Gaussian assumption for the errors is
sufficient. (2) For the different halo terms (2-halo term etc.), there
are uncertainties in the model, such as the halo exclusion
effect (Takada \& Jain 2003a; Fosalba et al. 2005). The
perturbation theory prediction is an approximate replacement for
these multiple-halo terms. Takada \& Jain (2003b) showed for example
that the the lensing three-point functions
computed in this manner are in better agreement with the simulations
than the standard halo model predictions (1-, 2- plus 3-halo terms).

Let us consider the contribution of the perturbation theory
trispectrum to the covariance. 
As derived in Appendix~\ref{app:bc}, the full PT mass
trispectrum (including the fundamental mode uncertainty) can be
computed as
\begin{eqnarray}
T^{\rm PT}_\delta(\bm{k}_a+\bm{\varepsilon}_{k}',
-\bm{k}_a+\bm{\varepsilon}_{k}^{\prime\prime},
\bm{k}_b+\tilde{\bm{\varepsilon}}_{k}^{\prime\prime\prime},
-\bm{k}_b+\tilde{\bm{\varepsilon}}_{k}^{\prime\prime\prime\prime})
&\approx&T^{\rm PT}_\delta(\bm{k}_a,-\bm{k}_a,\bm{k}_b,-\bm{k}_b)
\nonumber\\
&&+
8P^L_\delta(k_a)P^L_\delta(k_b)P^L_\delta(\varepsilon_{k})
F_2(\bm{\varepsilon}_{k},-\bm{k}_a)
\left[F_2(\bm{\varepsilon}_{k},\bm{k}_b) + F_2(\bm{\varepsilon}_{k},-\bm{k}_b)
\right],
\label{eqn:pttrisp_bc}
\end{eqnarray}
where $\bm{\varepsilon}_{k}$ with prime superscripts denote the
fundamental modes for a given survey and $\bm{\varepsilon}_k\equiv
\bm{\varepsilon}_k^\prime+\bm{\varepsilon}_k^{\prime\prime}$. The function
$F_2$ is the Fourier-space kernel of the 2nd-order
density perturbation defined by Eqn.~(\ref{eqn:f2}),
$P_\delta^L(k)$ is the linear-order mass power spectrum and
$T_\delta^{\rm PT}$ is the tree-level PT mass trispectrum. In
perturbation theory, $T_\delta^{\rm PT}\sim O([P^L_\delta(k)]^3)$ where
$k\sim k_a, k_b$, while the beat-coupling
term is of order $P^L_\delta(k)^2P^L_\delta(\varepsilon_k)$. So the latter 
is greater than the former if $P^L_\delta(\varepsilon_{k}) > 
P_\delta^L(k)$, which holds for modes of interest for the 
CDM power spectrum on wavenumber larger than the turnover scale, 
which are accessible from current and upcoming surveys.

In the ``translinear'' regime where the beat-coupling
contribution is most significant in the covariance, 
it roughly gives ${\rm Cov}\sim P(\varepsilon)P(k)P(k')$ because the
kernel $F_2\sim O(1)$ or equivalently the quantity ${\rm
Cov}/P(k)P(k')\sim P(\varepsilon )={\rm constant}$. This is consistent
with the plateau seen for $k\simgt 0.2$ in Fig.~2 in Hamilton et
al. (2006).  In this regime, the 
the signal-to-noise ratio for
the power spectrum amplitude (defined below in 
\S~\ref{sec:sn}) ceases to grow with increasing wavenumber 
(also see Rimes \& Hamilton 2005 and
Neyrinck et al. 2006 for simulation- and halo model based studies, 
and Lee \& Pen 2008 measurements from the SDSS galaxy
power spectrum). In the next section we will show the
results for lensing, which are modified by 
line-of-sight projections. 

Substituting Eqn.~(\ref{eqn:pttrisp_bc}) into Eqn.~(\ref{eqn:pscov})
gives the contribution to the covariance. 
The angle integration for the beat-coupling terms,
combined with the Limber's approximation, is given by 
\begin{equation}
\oint\frac{d\theta_{\bm{k}}}{2\pi }F_2(\bm{\varepsilon}_{k},
\bm{k})=\frac{6}{7},  
\end{equation}
so that that the PT trispectrum contribution to the non-Gaussian
errors of the lensing covariance can be expressed as
\begin{eqnarray}
{\rm Cov}[P_{(ij)}(l_a),P_{(i'j')}(l_b)]^{\rm NG, PT}
&\approx& \frac{1}{4\pi f_{\rm sky}}\left[
\int^{2\pi}_0\!\!\frac{d\theta}{2\pi} T^{\rm
PT}_{(iji'j')}(l_a,l_b,\cos\theta)\right.
\nonumber\\
&&\hspace{-3em}\left.+16\left(\frac{6}{7}\right)^2
\int\!\!d\chi~ W_{(i)}W_{(j)}W_{(i')}W_{(j')}
\chi^{-6}P_\delta^L\!\left(k_a=\frac{l_a}{\chi}\right)
P_\delta^L\!\left(k_b=\frac{l_b}{\chi}\right)
P_\delta^L\!\left(\varepsilon_{k}=\frac{2\pi}{L\chi}\right)
\right],
\label{eqn:lensng_pt}
\end{eqnarray}
where 
\begin{equation}
T^{\rm PT}_{(iji'j')}(l_a,l_b,\cos\theta)
=\int\!\!d\chi~ W_{(i)}W_{(j)}W_{(i')}W_{(j')}
\chi^{-6}T^{\rm PT}_\delta\left(\bm{k}_a,-\bm{k}_a,\bm{k}_b,-\bm{k}_b;
\cos\theta\right),
\end{equation}
and $\bm{k}_a\cdot\bm{k}_b\equiv k_ak_b\cos\theta$. Note that the
covariance above depends on the survey size through the prefactor
$f_{\rm sky}=L^2/4\pi=\Omega_{\rm s}/4\pi$ and the additional 
dependence of the beat-coupling term on $\varepsilon_{k}$.

Next let us consider the 1-halo term contribution, which dominates at
small angular scales.  
Although the trispectrum generally depends on four wavevectors,  
the 1-halo term depends only on the length of each vector since 
we assume spherical halos. 
This allows us to make the approximation,
even in the presence of the beat-coupling 
contamination, as
\begin{eqnarray}
{\rm Cov}[P_{(ij)}(l_a),P_{(i'j')}(l_b)]^{{\rm NG}, 1h}&=&\frac{1}{4\pi
 f_{\rm sky}}\int_{|\bm{l}|\in l_a}\!\!\frac{d^2\bm{l}}{A(l_a)}
\int_{|\bm{l}'|\in l_b}\!\!\frac{d^2\bm{l}'}{A(l_b)}
T_{(iji'j')}^{1h}(|\bm{l}+\bm{\varepsilon}|
,|-\bm{l}+\bm{\varepsilon}'|,|\bm{l}'+\bm{\varepsilon}^{\prime\prime}|
,|-\bm{l}'+\bm{\varepsilon}^{\prime\prime\prime}|)\nonumber\\
&\simeq&\frac{1}{4\pi
 f_{\rm sky}}\int_{|\bm{l}|\in l_a}\!\!\frac{d^2\bm{l}}{A(l_a)}
\int_{|\bm{l}'|\in l_b}\!\!\frac{d^2\bm{l}'}{A(l_b)}
T_{(iji'j')}^{1h}(l,l,l',l')\nonumber\\
&=&\frac{1}{4\pi f_{\rm sky}}T_{(iji'j')}^{1h}(l_a,l_a,l_b,l_b),
\label{eqn:lensng_1h}
\end{eqnarray}
where  we have 
assumed that the multipoles of interest, where the non-Gaussian errors
are relevant, are greater than the fundamental mode of a given survey:
$l_a,l_b\gg \varepsilon=2\pi/L$, and on the third equality of the
r.h.s. we have assume the trispectrum varies smoothly within bins of
$l$. 
Thus the
1-halo term is not affected by the beat-coupling
contamination. 

For a given model, we compute the lensing covariance as the sum of 
the Gaussian contribution, the first term on the r.h.s. of
Eqn.~(\ref{eqn:pscov}), and the non-Gaussian contributions, given by
${\rm Cov}^{\rm NG,PT}$ in Eqn.~(\ref{eqn:lensng_pt}) and 
${\rm Cov}^{{\rm NG,}1h} $ in Eqn.~(\ref{eqn:lensng_1h}). 
As one demonstration of our halo model approach we compare the model
predictions for the covariances of angular galaxy power spectrum with
the SDSS measurement result of Lee \& Pen (2008) in
Appendix~\ref{app:leepen}, where an encouraging agreement is found.

\section{Results}
\label{sec:results}
\subsection{Model Parameters}
\label{sec:modelparas}

To compute the lensing observables, we need to specify a cosmological
model and survey parameters. 
We will forecast below how the non-Gaussian errors degrade
parameter determination through measurements of power spectra as a 
function of survey parameters that are chosen to represent future weak
lensing surveys (from both ground and space). 

We include all the key parameters that may affect lensing observables
within the CDM and dark energy cosmological framework. Our fiducial
model is based on the WMAP 5-year results (Komatsu et al. 2008):
the density parameters for dark energy, CDM and baryon are $\Omega_{\rm
de}(=0.74)$, $\Omega_{\rm cdm}h^2(=0.11)$, and $\Omega_{\rm
b}h^2(=0.0227)$ (note that we assume a flat universe); the
primordial power spectrum parameters are the spectral tilt,
$n_s(=0.963)$, the running index, $\alpha_s(=0)$, and the normalization
parameter of primordial curvature perturbations, $A_s\equiv
\delta_\zeta^2(=2.41\times 10^{-9})$ (the values in the parentheses
denote the fiducial model). We employ the transfer function of matter
perturbations, $T(k)$, with baryon oscillations smoothed out (Eisenstein
\& Hu 1999), and adopt the primordial power spectrum given in Appendix C
in Takada et al. (2006), where the primordial spectrum amplitude is
normalized at $k_0=0.002~$Mpc$^{-1}$ following the convention in Komatsu
et al. (2008).  Note that the present-day rms mass fluctuations enclosed
within spheres of radius $8h^{-1}$Mpc, $\sigma_8\simeq 0.80$ for our
fiducial model.
The dark energy equation of state, which governs redshift evolution of
the energy density of dark energy together with $\Omega_{\rm de}$, is
parametrized as $w(a)=w_0+w_a(1-a)$, with fiducial values $w_0=-1$
and $w_a=0$.

We model the redshift distribution of galaxies with a function specified by
one parameter $z_0$ (which depends on survey depth):
\begin{equation}
n_s(z;z_0)=n_0 \times 4z^2 \exp\left[-\frac{z}{z_0}\right],
\label{eqn:ns}
\end{equation}
where the normalization is fixed by setting $n_0=1.18\times 10^9$ per unit
steradian corresponding to $n_0\simeq 100 $ arcmin$^{-2}$. The redshift
dependence is same as that assumed in Takada \& Jain (2004) or Huterer
et al. (2006), but the normalization is fixed.  This simple
form has nice properties as follows. The mean redshift of source galaxies,
$z_m$, is given by $z_m \equiv \int^\infty_0dz z n_s(z; z_0)/
\int^\infty_0dz n_s(z; z_0)=3z_0$. Thus once the mean redshift is
specified, the average number density of galaxies (or equivalently the
survey depth) is also specified: e.g., for $z_m=0.7, 1.0, 1.2$ and $1.5$
the corresponding number densities are $\bar{n}_g\equiv \int_0^{\infty}\!dz~
n(z)\simeq 10, 30, 51$ and $100$ arcmin$^{-2}$, respectively. Hence the
source distribution above can roughly represent future planned lensing
surveys, DES, Subaru Weak Lensing Survey, LSST, and SNAP, simply by
adjusting the parameter $z_0$ (e.g. Hoekstra \& Jain 2008 for a review).

The fiducial survey we will assume in the following is a ground-based
weak lensing survey that is given by survey area $2000$ deg$^2$, mean
source redshift $z_m=1$ corresponding to $\bar{n}_g\simeq 30$
arcmin$^{-2}$, and rms intrinsic ellipticity
$\sigma_{\epsilon}=0.22$.  
Our fiducial survey roughly resembles the Subaru Weak Lensing
Survey (Miyazaki et al. 2006).

\subsection{Correlation Coefficients of the Power Spectrum Covariance}

The correlation coefficients of the power spectrum covariances quantify
the relative strengths of the off-diagonal components to the diagonal
components; e.g., the correlation strengths between band powers at
different multipole bins for the case of no tomography.  The correlation
coefficient is defined from Eqn.~(\ref{eqn:pscov}) as
\begin{equation}
r_{AB}\equiv \frac{\bm{C}_{AB}}{\sqrt{\bm{C}_{AA}{\bm C}^{\rm
 }_{BB}}}.  \label{eqn:rpij}
\end{equation}
The coefficients are normalized so that $r=1$ for the diagonal
components with $A=B$.  For the off-diagonal components with $A\ne B$,
$r_{AB}\rightarrow 1$ implies strong correlation between the power
spectra of $A$- and $B$-th bins, while $r_{AB}=0$ corresponds to no
correlation.  As shown in Eqn.~(\ref{eqn:pscov}), the relative importance
of the non-Gaussian errors in the power spectrum covariance depends on
the bin width of angular multipoles assumed. For larger bin width, 
the non-Gaussian error contributions get suppressed relative to
the Gaussian errors.

\begin{figure}
  \begin{center}
    \leavevmode\epsfxsize=12.cm \epsfbox{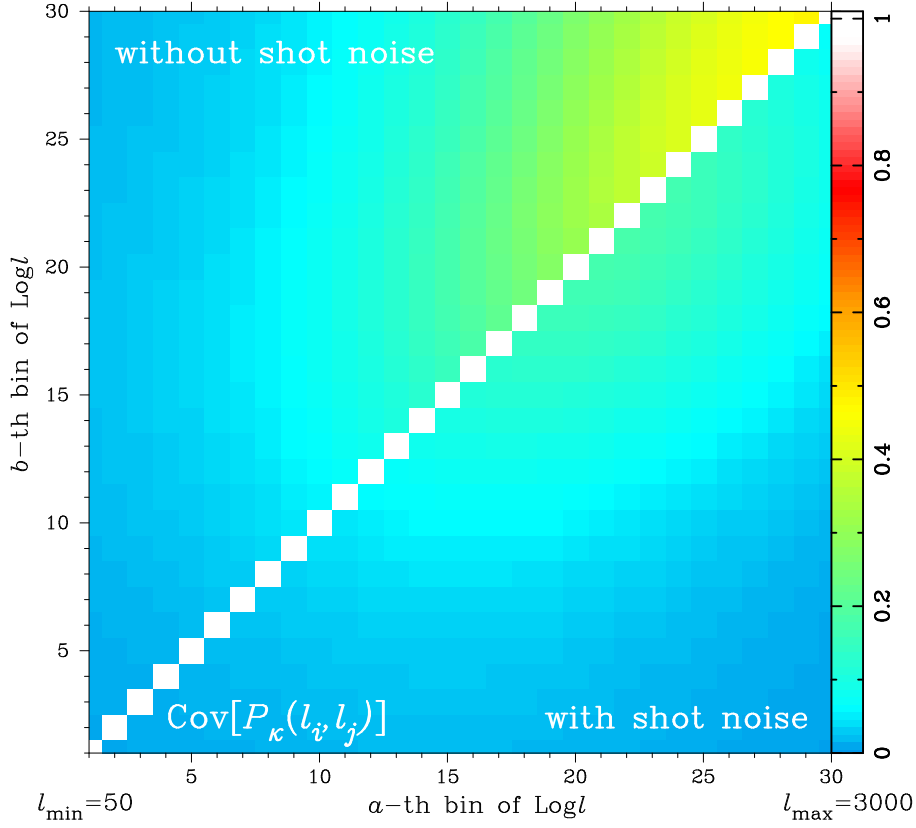}
  \end{center}
\caption{Power spectrum cross-correlation coefficients $r_{AB}$ (see
Eqn.~[\ref{eqn:rpij}]) for our fiducial ground-based lensing survey.
We consider a
single redshift bin (no tomography case), so the indices $(A,B)$
run over $30$ logarithmic bins in multipole space over the
range $50\le l\le 3000$. The off-diagonal components arise
purely from the non-Gaussian errors as shown in
Eqn.~(\ref{eqn:pscov}). 
The upper-left and
lower-right off-diagonal components are the results without and with
shot noise contamination due to intrinsic ellipticities, respectively.
One can see significant correlations at higher $l$.  
However, shot noise suppresses the relative importance
of non-Gaussian correlations.  
 } \label{fig:rpij}
\end{figure}
Figure~\ref{fig:rpij} shows the correlation coefficients for the
fiducial ground-based survey above and the case of no tomography.
One can see that the correlation coefficients are well below 0.5 for
low multipoles, but increase as one goes to smaller angular scales,
because the lensing signal is more affected by non-linear clustering. 
This result can be compared to the coefficients of the 3D mass power
spectrum, where strong correlations, $r\simgt 0.5$, can be seen even in
the weakly non-linear regime (Scoccimarro et al. 1999; Takahashi et
al. in prep.). The weaker
correlations in the lensing power spectrum are due to the line-of-sight
projections of independent lensing structures at different redshifts,
making the weak lensing fields closer to the Gaussian limit. 

Although the covariance matrix is symmetric, we have chosen to 
show different cases in the upper-left and lower-right elements: 
intrinsic ellipticity contributions to the covariance are included
only in the lower-right.  Because the
shot noise only contributes to the Gaussian terms (diagonal elements) in
the covariance matrix (the denominator of Eqn.~[\ref{eqn:rpij}] for
$r_{AB}$), it lowers $r$. As one considers smaller angular
scales where the multipoles lie well in the shot-noise dominated regime,
the power spectrum becomes less affected by the non-Gaussian errors for
a given survey. Therefore, a survey with
smaller number density of source galaxies is relatively less affected by
the non-Gaussian errors.  

We have also checked that our model predictions fairly well reproduce
the simulation results for the correlation coefficients shown in Table~2
in Cooray \& Hu (2001; also see White \& Hu 2000) for multipoles 
$200\simlt l\simlt 2000$ (when we adjusted our model parameters
and multipole bin widths to match those used by these authors).

\subsection{Signal-to-Noise for the Lensing Power Spectrum}
\label{sec:sn}

\begin{figure*}
  \begin{center}
    \leavevmode\epsfxsize=13.cm \epsfbox{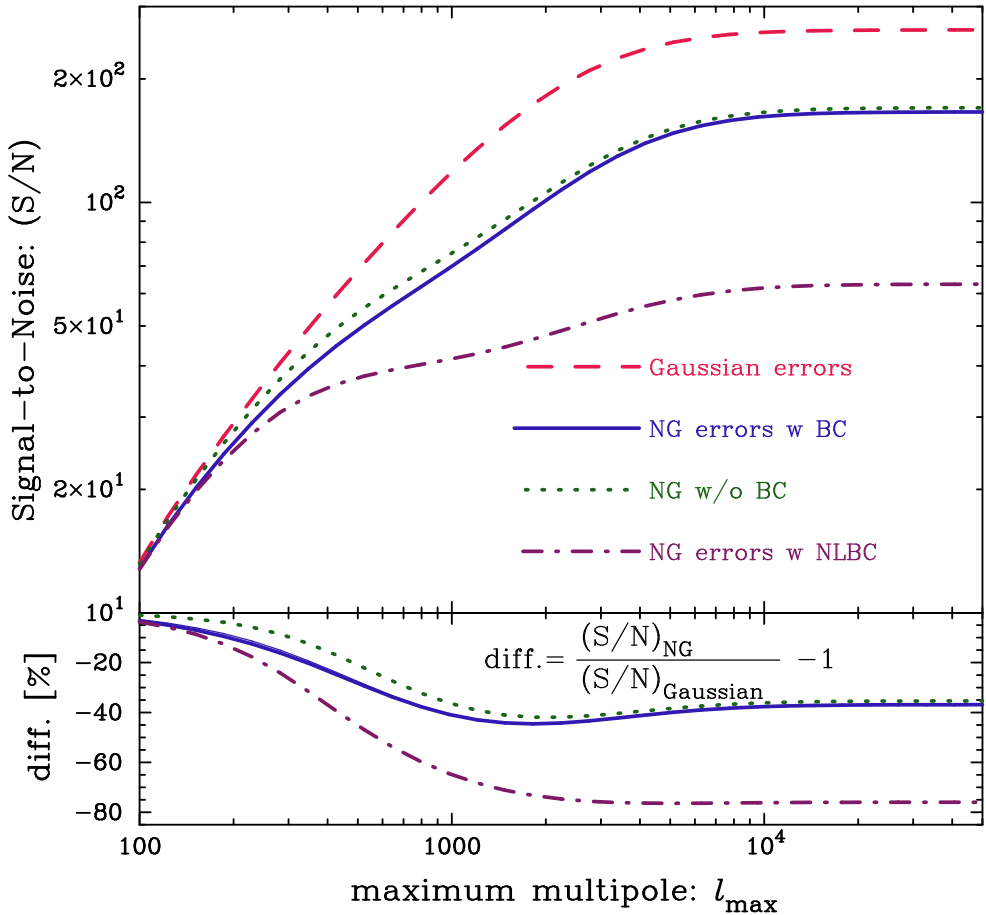}
  \end{center}
\caption{The expected cumulative 
signal-to-noise ratio ($S/N$)
 for the power spectrum is shown 
as a function of maximum multipole $l_{\rm
 max}$, where the power spectrum information over $50\le l\le l_{\rm
 max}$ is included. The dashed and bold-solid curves show the results
without and with the non-Gaussian error contribution 
to the covariance,  
 respectively. The dotted curve shows the result when the
 beat-coupling contribution is ignored, while the dot-dashed curve is
 the result when the nonlinear beat-coupling effect is assumed (see text
 for the details). The survey parameters are as
 in Figure~\ref{fig:ps}. 
The lower panel shows the percentage difference
relative to the $S/N$ with Gaussian errors. 
}
 \label{fig:sn_lmax}
\end{figure*}
A useful way to quantify the impact of the non-Gaussian errors is to
study the expected signal-to-noise ratio ($S/N$) for measuring the
lensing power spectrum from a given survey, which is independent of the
multipole bin widths assumed, as long as the power spectrum does not vary
rapidly within the bin widths. 
The $S/N$ may be defined, using the covariance from 
Eqn.~(\ref{eqn:pscov}), as
\begin{equation}
\left(\frac{S}{N}\right)^2=
\sum_{A,B} P_{(ij)}(l_a) \left[\bm{C}^{-1}\right]_{AB}P_{(i'j')}(l_b),
\label{eqn:snps1}
\end{equation}
where $\bm{C}^{-1}$ is the inverse of the covariance matrix and the
summation indices $A,B$ correspond to the dimension of the covariance
matrix and run over multipole bins and redshift bins.  Note that the
power spectra in the above equation represent the lensing signal and
do not include the shot noise.
The $S/N$ defined above is equivalent to the Fisher information content
studied in Tegmark et al. (1997), Rimes \& Hamilton (2005) and Lee \&
Pen (2008). 
It describes the accuracy in measuring the
amplitude of the lensing power spectrum when the shape is completely
known.

Figure~\ref{fig:sn_lmax} shows the expected
$S/N$ as a function of the maximum multipole $l_{\rm max}$, where the
power spectrum information over $50\le l \le l_{\rm max}$ is included in
the $S/N$ calculation. 
The solid curve
shows the $S/N$ obtained from our fiducial model prediction for the
lensing covariance developed in \S~{\ref{sec:cov}}. It can be
compared with the result without the non-Gaussian sample variances,
i.e. for the Gaussian error case. 
As expected, the $S/N$ increases with increasing $l_{\rm max}$ due to the
gain in multipole modes probed. However, the $S/N$ ceases to increase at
$l_{\rm max}\simgt 4000$, i.e. little cosmological 
information is available from
these high multipoles, because the shot-noise contamination becomes dominant
in the covariances. The impact of the non-Gaussian errors on the $S/N$
also changes with $l_{\rm max}$ in a characteristic way.
For small $l_{\rm max}$ such as $l_{\rm max}\simlt 500$, the effect is
small, i.e. the $S/N$ is close to the Gaussian case. For larger $l_{\rm
max}$ where the lensing fields are more affected by the nonlinear regime
of mass clustering, the $S/N$ decreases by up to a factor of 2 as
explicitly shown in the lower panel.  The dotted curve shows the result
obtained when the beat-coupling effect on the covariance due to a finite
survey area is ignored. This contribution
appears to be non-negligible over a range of $l_{\rm max}$, from a few
hundreds to $\sim 1000$. 
However, in contrast to the case of 3D mass power spectrum (Rime \&
Hamilton 2005; Neyrinck et al. 2006), 
a clear plateau shape in the $S/N$ curves on scales before the shot noise
dominated regime cannot
be seen. Hence the line-of-sight projection
appears to weaken 
the impact of non-Gaussian errors;  lensing
at a given angular scale arises from mass fluctuations over a
wide range of length scales that may span linear to
nonlinear regimes.

Since the magnitude of the beat-coupling effect is not yet well tested 
from simulations, we also estimate the ``worst-case'' impact 
by replacing the linear mass power spectra appearing
in Eqn.~(\ref{eqn:lensng_pt}) with the {\em nonlinear} power spectra.
This case, abbreviated NLBC for nonlinear beat coupling, 
is shown by the dot-dashed curve in Figure~\ref{fig:sn_lmax}.
Physically this may arise if the mass
distribution on highly nonlinear scales, such as the 
mass distribution within
a dark matter halo, is correlated with the mass fluctuations on very
large length scales (such correlations are in fact likely to be
very weak). The $S/N$ in this case saturates or increases very slowly 
for $l_{\rm max}\simgt 300$,
resulting in a substantial decrease in the $S/N$ (up to a factor of 5). 

In Appendix~\ref{app:leepen} we compare our halo model predictions
with the measurements by Lee \& Pen (2008) of the 
$S/N$ for the angular power spectrum of SDSS galaxies. 
The standard halo model developed in
\S~\ref{sec:cov} is in reasonable agreement with the measurements, 
consistent with the 10-30\% level agreement we found previously 
between the halo model and simulations (Takada \& Jain 2002,2003b).
Even so, we will continue to use the NLBC model 
to estimate the worst case impact of the non-Gaussian errors.  
Further detailed studies of the power spectrum covariances for 3D
mass and 2D lensing are in preparation (Takahashi et al. and Sato et
al. in preparation; also see Pielorz 2008). 

We also estimate the contribution made by the PT
trispectrum to the covariance: in Figure~\ref{fig:sn_lmax}, the
thin-solid curve in the lower panel shows the result when 
its contribution is ignored. 
The difference between the bold-
and thin-solid curves is barely visible (at $l_{\rm max}\sim
100$), therefore the PT trispectrum contribution is very small.

\begin{figure*}
  \begin{center}
    \leavevmode\epsfxsize=13.cm \epsfbox{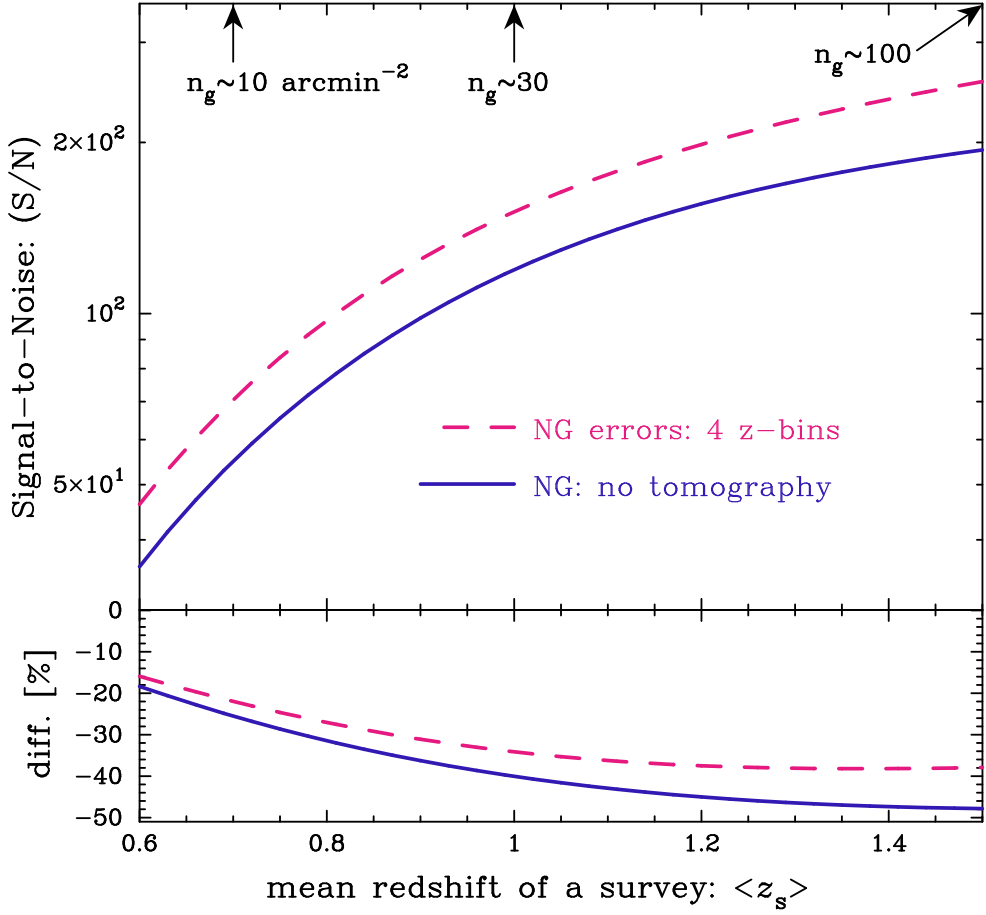}
  \end{center}
\caption{The solid curve shows the dependence of $S/N$ on 
mean source redshift for the source galaxy distribution given by 
Eqn.~(\ref{eqn:ns}), assuming  non-Gaussian errors. Power
 spectrum information over $50\le l \le 3000$ is included, and the
 survey parameters are as in Figure~\ref{fig:ps}. 
The $x$-axis label above the panel shows the 
corresponding number density of source galaxies. 
Deeper surveys are more affected by non-Gaussian errors, due to the 
reduced contribution of shot noise.
The dashed curve shows the
result for 4 $z$-bin redshift tomography. 
The lower panel shows the percentage degradation in $S/N$ compared 
to the Gaussian errors (with the same number of redshift bins in each
case). 
Adding redshift information not only 
increases the $S/N$, but also reduces the impact of the non-Gaussian
errors. 
}  \label{fig:sn_zm}
\end{figure*}
The solid curve in Figure~\ref{fig:sn_zm} 
shows how the $S/N$ changes with survey
depth, i.e. the mean source redshift $z_m\equiv \skaco{z_s}$.   Given a
mean source redshift the average number density of source
galaxies is specified by Eqn.~(\ref{eqn:ns}), 
which in turn determines the shot noise contribution to
the covariances. 
The $S/N$ is greater for a deeper survey, and increases 
by a factor of 5 between $z_m=0.6$ and $z_m=1.5$, because the lensing
signal is higher.  However, the impact of non-Gaussian errors on the $S/N$ also
becomes more significant for a deeper survey. 
These results are insensitive 
to other choices of $l_{\rm max}$ such as $l_{\rm
max}=1000$ or $10^4$, as shown in Figure~\ref{fig:sn_sig8}. 

The dashed curve in Figure~\ref{fig:sn_zm} demonstrates that adding
tomographic redshift information into the lensing power spectrum
measurement not only increases the total $S/N$'s (also see Fig.~5 in
Takada \& Jain 2004), but also reduces the impact of the non-Gaussian
errors. 
Compared to no tomography case, adding four redshift bins reduces the
impact of non-Gaussian errors on the $S/N$, e.g. to $\sim 35\%$ from
$\sim 50\%$ when $l_{\rm max}=3000$.  Note that four redshift bin
tomography is our fiducial survey design for our forecasts of parameter
determination shown below. 

Finally, in Appendix~\ref{app:sn} we show how the $S/N$ depends
on the shot noise and the $\sigma_8$ assumed. 

\subsection{Principal Component Analysis of Lensing Covariance}
\label{sec:pca}

\begin{figure}
  \begin{center}
    \leavevmode\epsfxsize=17.5cm \epsfbox{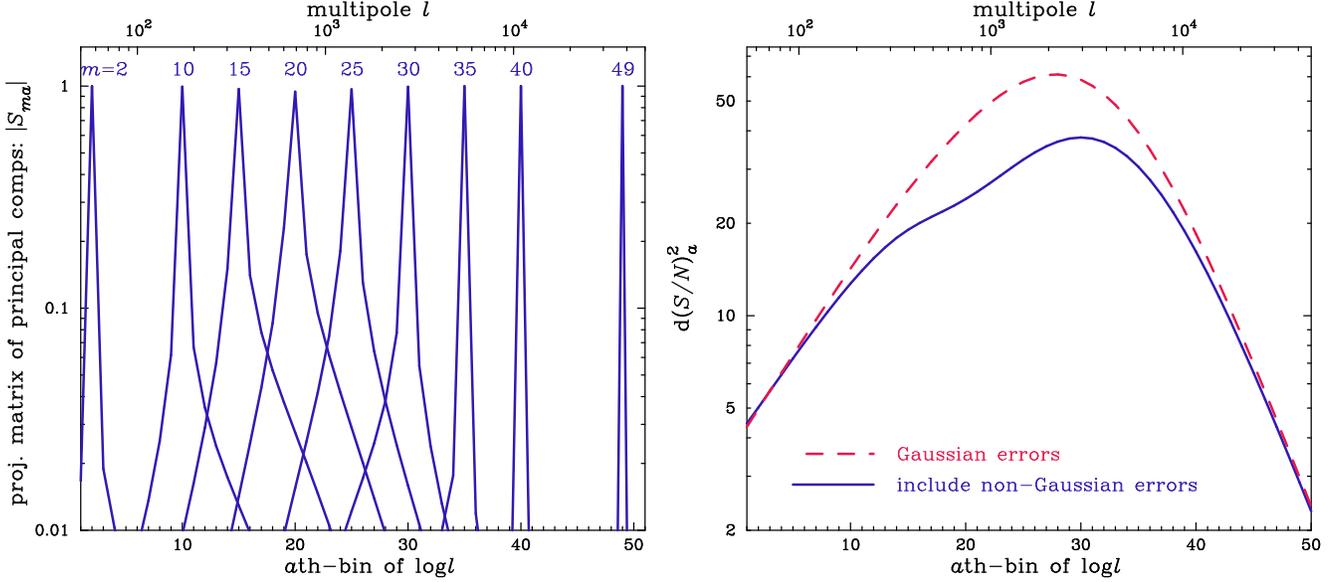}
  \end{center}
\caption{{\em Left panel}: The projection matrix of $S_{am}$ for the
 principal component decomposition of the power spectrum covariance is
 plotted as a function of multipole bins $l_a$.  
We employed 50 bins for the multipole binning, i.e. the
 dimension of $S_{am}$ is $50\times 50$, and we assumed the survey
 parameters same as in Figure~\ref{fig:ps}. The absolute value of
 $|S_{am}|$ for the eight cases of $m=2, 10, 15, 20, 25, 30, 35, 40, 49$ are
 shown.  The projection matrix around $l\sim 10^3$ gets 
 contributions from a wide range of neighboring multipoles. 
 {\em Right panel}: The differential contribution to the
 $(S/N)^2$ at each multipole bin. The dashed and solid curves show the
 results with the Gaussian and non-Gaussian errors, respectively. The
 power spectra with $l\sim 2000$ contribute most to the $S/N$, but are
 also strongly affected by the non-Gaussian errors. For a survey of 2000
 degree$^2$, $S/N \simgt 3$ can be expected in each multipole
 bins for $100\simlt l\simlt 2\times 10^4 $. The
 $d(S/N)^2$ plotted here scales with sky coverage roughly
 as $d(S/N)^2\propto f_{\rm sky}$.  }  \label{fig:pspca}
\end{figure}
A principal component analysis of the power spectrum covariance is
useful to quantify how the spectra of different multipoles are
correlated and how many independent modes exist
(see also Scoccimarro et al. 1999 for a similar study for the 3D mass power
spectrum). Since the covariance matrix is symmetric by definition, 
it can always be decomposed as
\begin{equation}
C_{ab}=\sum_{m} S_{am}(\lambda_m)^2S_{bm},
\end{equation}
where $\lambda_m$ is the $m$-th eigenvalue or principal component,
$\bm{S}=\bm{S}^{\rm T}$, $\sum_c S_{ac}S_{bc}=\delta_{ab}$ and $\bm{S}$
is normalized so as to satisfy $\sum_{b}(S_{ab})^2=1$. 
We consider here no tomography case for
simplicity, therefore the dimension of the covariance is given by number of
multipole bins.  The matrix $S_{am}$ is considered as the projection
matrix as it describes how the power in the $a$-th multipole bin is projected
onto the $m$-th eigenmode.  Using this representation, the inverse of
the covariance matrix is given by
$[\bm{C^{-1}}]_{ab}=\sum_{m}S_{am}(1/\lambda_m)^2S_{bm}$. Hence, the
signal-to-noise given by Eqn.~(\ref{eqn:snps1}) can be rewritten
as
\begin{equation}
\left(\frac{S}{N}\right)^2=
\sum_{m}\left\{ 
\frac{1}{\lambda_m}\sum_{a} S^{\rm P}_{am}P_a
\right\}^2.
\label{eqn:snps}
\end{equation}
The equation above expresses the $S/N$ as a sum
of contributions from independent eigenmodes.

The projection matrix elements, $|S_{am}|$, for
selected values of $m$ between 2 and 50
are shown against the multipole bins in the left
panel of Figure~\ref{fig:pspca}.  Note that elements of $S_{am}$
can be both positive and negative. 
For the $m$-th element, 
$|S_{am}|$ peaks at $a=m$. 
The projection matrix $S_{am}$ quantifies the correlation
between band powers of neighboring bins. 
The horizontal error bars in
Figure~\ref{fig:ps} are defined as the range of $|S_{am}|\ge 0.1$ around
each multipole bins. 
The projection matrices for
the modes around $l\sim 10^3$ are found to have broader tails,
reflecting stronger correlations between neighboring multipole bins due
to the non-Gaussian errors. For the modes of $l\sim 100$ or $l\simgt
4000$, $S_{am}$ has a steep peak at $a=m$, i.e. it is close to the
diagonal matrix. At the high $l$ end this is due to the dominance of
shot noise over non-Gaussian terms in the covariance. 
Using this approach, a 
more sophisticated decorrelating scheme of the
band powers may be developed extending the method for the galaxy power
spectrum measurement  (e.g., Tegmark et al. 2002) to the lensing case.

The right panel of Figure~\ref{fig:pspca} shows the differential
contributions of each principal component to the total $S/N$
for the fiducial survey. 
Note that this plot is shown as a function of the
multipole bins rather than the principal components, assuming that the
$m$-th principal component arise mainly from the $m$-th multipole bin as
shown in the left panel. The solid and dashed curves display the results
when the non-Gaussian errors are included or ignored, respectively.  The
power spectrum at $l\sim 3000$ is most accurately measurable for both
cases. 
More generally, a detection of the lensing signal 
at more than $3\sigma$ can be expected over the wide range $100\simlt
l\simlt 10^4$
even in the
presence of non-Gaussian errors. It should be noted that $(S/N)^2
\propto f_{\rm sky}$ (for a fixed $\bar{n}_g$) to good approximation,
therefore the result shown here can be scaled to a survey of
arbitrary area. Non-Gaussian errors degrade the
differential $S/N$ by up to a factor of 2.
The vertical error bars at each multipole bin in Figure~\ref{fig:ps}
are computed using: $\sigma(P(l_m))/P(l_m)=\pm [d(S/N)^2_m]^{-1/2}$.

We may summarize the results so far: 
non-Gaussian errors do affect the lensing power spectrum
measurement, and therefore must be included in measurement
analyses for ongoing and planned future
surveys (see also Semboloni et al. 2007). 
The impact of non-Gaussian errors depends on various ingredients
such as survey 
parameters, range of multipoles and cosmological parameters (especially
$\sigma_8$), and these dependencies need to be carefully taken into
account if a calibration of the covariance is
done based on simulations which may not span the full range of 
parameters.

\subsection{The Impact of Non-Gaussian Errors on Parameter Estimations}
%
\begin{table}
\begin{center}
Cosmological Parameters: Lensing+CMB\\ 
\begin{tabular}
{l|\colskip l\colskip l\colskip l\colskip l} \hline\hline
& Gaussian & Non-Gauss. 
& Non-Gauss. w/o BC & Non-Gauss. w NLBC \\
\hline
$\sigma(\Omega_{\rm de})$& 0.018 &0.019(6\%)&0.019(6)&0.021(17)\\
$\sigma(\ln A_s)$& 0.024&0.025(4\%)&0.024(0)&0.024(0)\\
$\sigma(w_0)$&0.21 & 0.22(5\%)&0.22(5)&0.24(14)\\
$\sigma(w_a)$&0.61 & 0.63(3\%)&0.62(2)&0.67(10)\\
$\sigma(n_s)$&0.012&0.012(0\%)&0.012(0)&0.012(0)\\
$\sigma(\alpha_s)$&0.0033&0.0034(3\%)&0.0033(0)&0.0034(3)\\ \hline
$\sigma(w_{\rm pivot})$&0.056 &0.061(9\%)&0.058(4)&0.065(16) \\
\hline\hline
\end{tabular}
\end{center}
\caption{Summary of parameter constraints for lensing
 tomography with 4 redshift bins, combined with  Planck CMB
 priors. Column 2: assuming Gaussian errors; 
Column 3: including the non-Gaussian errors; Columns
 4 and 5: the errors for the non-Gaussian errors, but ignoring the
 beat-coupling contribution, and assuming the non-linear beat-coupling
 (NLBC) contribution to the non-Gaussian errors (see text for the
 details), respectively. The numbers in round bracket show degradation
 in the errors compared to the Gaussian case. All errors are 68\%
 confidence-level errors and include marginalization over other
 parameters. Note that we include power spectrum information over the
 multipole range $50\le l \le 3000$ and assuming fiducial survey
 parameters as in Figure~\ref{fig:ps}: $\Omega_{\rm survey}=2000$
 degree$^2$, $\bar{n}_g\simeq 30$ arcmin$^{-2}$, $\skaco{z_s}\simeq 1$,
 and $\sigma_{\epsilon}=0.22$. } \label{tab:paras}
\end{table}
%

In this section we address how the non-Gaussian errors degrade the
ability of a given weak lensing survey to constrain cosmological
parameters, especially the parameters of dark energy. 

To do this, we use the Fisher matrix formalism to estimate accuracies of
estimating parameters given the power spectrum measurement.  
The parameter forecasts we obtain
depend on the fiducial model and are also sensitive to the choice of
free parameters. Our fiducial parameters of the cosmological model and
lensing survey are given in \S~\ref{sec:modelparas}.

Using the covariance matrix (\ref{eqn:pscov}), the Fisher matrix is
given by
\begin{equation}
\bm{F}^{\rm WL}_{\alpha\beta}=\sum_{A,B} \frac{\partial
 P_{(ij)}(l_a)}{\partial p_\alpha} \left[\bm{C}^{-1}\right]_{AB}
 \frac{\partial P_{(i'j')}(l_b)}{\partial p_\beta},
\end{equation}
where $p_\alpha$ ($\alpha=1,2,\dots$) denotes a set of parameters. The
marginalized 1$\sigma$ error on the $\alpha$-th parameter $p_\alpha$ is given
by $\sigma^2(p_\alpha)=(\bm{F}^{-1})_{\alpha\alpha}$, where
$\bm{F}^{-1}$ is the inverse of the Fisher matrix. It is sometimes
useful to consider projected constraints in a two-parameter subspace to
see how the two parameters are correlated, and this can be studied
following the method described in \S~4.1 in Takada \& Jain (2004).

Weak lensing alone cannot constrain all the cosmological parameters
simultaneously due to parameter degeneracies. However, the
parameter degeneracies will be efficiently broken by combining the weak
lensing constraints with constraints from the CMB temperature and
polarization anisotropies (e.g. Takada \& Jain
2004).  When computing the Fisher matrix for the CMB, we employ 9
parameters in total: the Thomson scattering optical depth to the last
scattering surface, $\tau(=0.087)$ plus the 8 parameters described in
\S~\ref{sec:modelparas}.  We use the publicly-available CAMB code (Lewis
et al. 2000), based on  CMBFAST (Seljak \&
Zaldarriaga 1996), to compute the angular power spectra of temperature
anisotropy, $C^{\rm TT}_l$, $E$-mode polarization, $C^{\rm EE}_l$, and
their cross correlation, $C^{\rm TE}_l$. Note that we ignore the
$B$-mode spectra arising from the primordial gravitational waves.
Specifically we consider the noise per pixel and the angular resolution
of the Planck experiment that were assumed in Eisenstein et
al. (1998). In this calculation we use the range of multipoles $10\le
l\le 1500$ for $C^{\rm TT}_l$ and $C^{\rm TE}_l$ and use $2\le l \le
1500$ for $C_l^{\rm EE}$, respectively. Therefore we do not include the
ISW effect on the temperature spectra at low multipoles $l\simlt 10$
which might be affected by a possible contribution of clustered dark energy.
To be conservative, however, we do not include the CMB information on
dark energy equation of state parameters, $w_0$ and $w_a$
(also see Takada \& Bridle 2007). We first compute the
inverse of the CMB Fisher matrix, $\bm{F}^{-1}_{\rm CMB}$, for the 9
parameters in order to obtain marginalized errors on the parameters, and
then re-invert a sub-matrix of the inverse Fisher matrix that includes
only the rows and columns for the parameters beside $w_0$ and $w_a$. The
sub-matrix of the CMB Fisher matrix derived in this way describes
accuracies of the 7 parameter determination, including degeneracies with
the dark energy parameters $w_0$ and $w_a$ for the
hypothetical Planck data sets.  
The CMB Fisher matrix is added to the lensing
Fisher matrix as $\bm{F}_{\alpha\beta}=\bm{F}^{\rm WL}_{\alpha\beta}
+\bm{F}^{\rm CMB, sub}_{\alpha\beta}$ ($\alpha,\beta=1,2,\dots,9$),
where the elements of the CMB Fisher sub-matrix including dark energy
parameters $w_0$ or/and $w_a$ are set to zero.

\begin{figure}
  \begin{center}
    \leavevmode\epsfxsize=12.cm \epsfbox{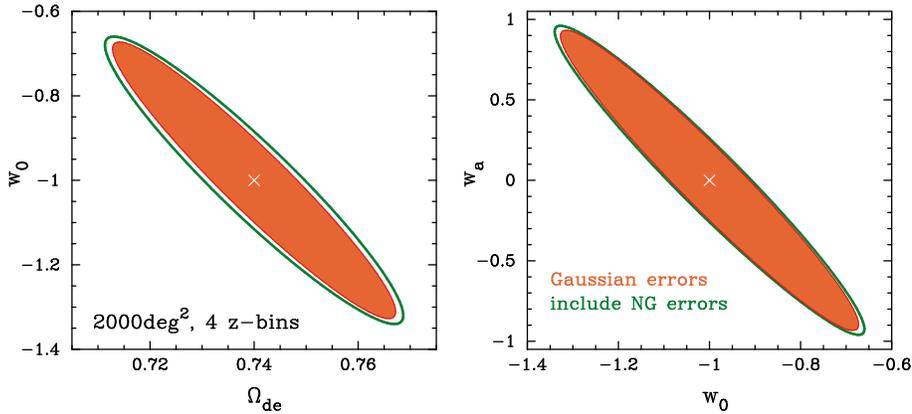}
  \end{center}
\caption{Fisher errors on dark energy parameters, 
with and without non-Gaussian errors in the lensing covariances.  
We use lensing tomography with 4 redshift bins and Planck CMB priors.
 The non-Gaussian errors enlarge the areas of the error ellipses in
the subspaces of ($\Omega_{\rm de}, w_0$) and ($w_0, w_a$) by 
$26\%$ and $12\%$, respectively. Table~\ref{tab:paras} gives the
one-dimensional marginalized errors on each parameter. 
}
\label{fig:chi2}
\end{figure}
%
\begin{figure}
  \begin{center}
    \leavevmode\epsfxsize=12.cm \epsfbox{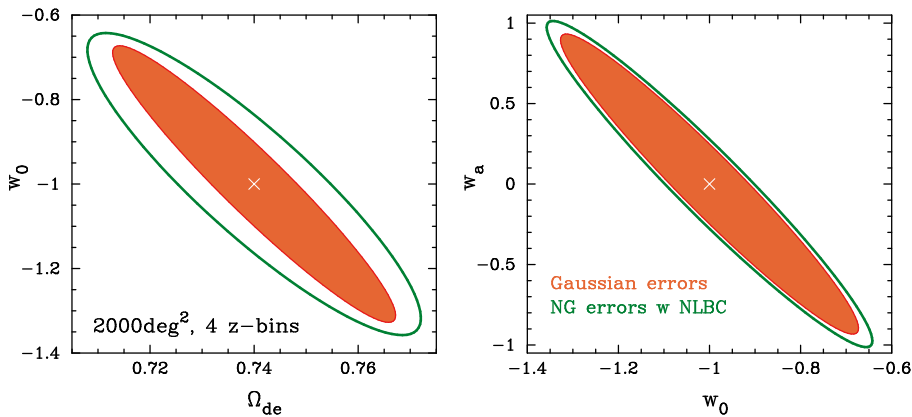}
  \end{center}
\caption{As in the previous figure, but using the non-linear
 beat-coupling contribution for the non-Gaussian errors. This
enlarges the error ellipses (compared to the Gaussian case) 
by a factor of 2 and  $\sim 25\%$, respectively. }
\label{fig:chi2_nlbc}
\end{figure}
Table~\ref{tab:paras} summarizes forecasts for parameter constraints
that are expected from lensing tomography of 4 redshift bins (with CMB
priors), for our fiducial survey
parameters as in Figure~\ref{fig:ps}.  
We include the
power spectrum information over  $50\le l \le 3000$
and bin the galaxies such that their number densities 
in each redshift bin are about equal for a given redshift 
distribution of galaxies
(see Eqn.[\ref{eqn:ns}]). Notice that $\sigma(w_{\rm
pivot})$ shows the error in the dark energy equation of state at the
best constrained redshift, {\em pivot redshift} -- it is equivalent to 
the error on a constant $w=w_0$, with $w_a$ fixed to the fiducial value.  

One of our main results is given in Column 3 in Table~\ref{tab:paras}, 
which gives parameter errors including non-Gaussian 
covariances. These degrade parameter errors typically by less than $10\%$. 
This is much smaller than the degradation in the
signal-to-noise ratio of power spectrum measurement shown in 
Figure~\ref{fig:sn_zm} (a factor of $0.65$).  
Note that the unmarginalized error on each parameter is degraded
by roughly similar amount to that for the $S/N$. 
If we take the volume of the Fisher error ellipsoid (in our
8-dimensional parameter space) as inversely proportional to 
the $S/N$ magnitude, then 
non-Gaussian errors enlarge the Fisher volume by a factor of
about $1.54\ (\simeq 1/0.65)$.  So if all the eight principal axes of
the Fisher ellipsoid are equally stretched by the non-Gaussian errors,
each parameter error would be degraded only by about $6\%[\simeq
(1.54)^{1/8}-1]$, which is the maximum degradation seen in 
Table~\ref{tab:paras}. In reality, lensing carries information from
distance factors, which is not degraded by  non-Gaussian errors. 
Also, non-Gaussian errors change the directions
of the principal axes, or equivalently
the directions and degrees of parameter degeneracies in parameter
space, and we have further combined with the CMB information -- both
effects can change the results for a particular parameter. 

Our results in Table~\ref{tab:paras} can be compared with Table 4 in
Cooray \& Hu (2001). They found about a $15\%$ increase in 
parameter errors due to non-Gaussian errors, with a set of 5 parameters
for the Fisher parameter forecasts. Their results can be understood in
terms of our findings as follows.  First, a fiducial cosmology with
higher normalization $\sigma_8\simeq 1$ was assumed in contrast to our
$\sigma_8\simeq 0.8$. They also considered no tomography
case. For these assumptions non-Gaussian errors degrades the $S/N$ by
a factor 2 (see Figure~\ref{fig:sn_sig8}). Further
since they worked with fewer parameters, 5
in contrast to our 8 parameters, the degradation in the
marginalized error of each parameter is expected to be $\sim 15\%\simeq
2^{1/5}-1$.  Therefore our results are
consistent with theirs. 

In summary the impact of non-Gaussian errors
on parameter estimations depends on the number of parameters used and the
priors used in the analysis. The degradation in the marginalized
accuracy of a parameter of interest is generally
smaller than that in the total $S/N$ of the power spectrum. 

Column 4 in Table~\ref{tab:paras} shows the errors when ignoring the
beat-coupling contribution to the non-Gaussian errors -- the effect 
is very small. Column 5 shows the results when using the
nonlinear beat-coupling contribution to the non-Gaussian errors as
studied in Figure~\ref{fig:sn_lmax}.   This model is intended to
be an upper bound on the beat-coupling contribution.
But even in this case non-Gaussian errors degrade parameter accuracies 
by less than $20\%$.  

The non-Gaussian errors induce correlations between band powers of the
lensing power spectra in multipole as well as redshift space.
One may expect that the
correlations degrade parameters that are more sensitive to the
amplitude of the lensing power spectrum as discussed in Cooray \& Hu
(2001) and Takada \& Bridle (2007).  
Table~\ref{tab:paras} does show
that the error on $\Omega_{\rm de}$ is more degraded than the
other parameters ($\Omega_{\rm de}$ is most sensitive to the lensing spectrum
amplitude for a flat universe, and is also better constrained by
lensing than the CMB). 

Figures~\ref{fig:chi2} and \ref{fig:chi2_nlbc} 
show how non-Gaussian errors enlarge
the projected error ellipses in two-parameter subspaces of dark energy
parameters.  Our fiducial model for non-Gaussian errors predicts that
error ellipses are only slightly enlarged, while including the nonlinear
beat-coupling effect enlarges the area of error ellipse by a factor of 2
in the $(\Omega_{\rm de}, w_0)$-subspace, and the area by $\sim 25\%$ in
the ($w_0,w_a$)-subspace.

\begin{figure}
  \begin{center}
    \leavevmode\epsfxsize=17.cm \epsfbox{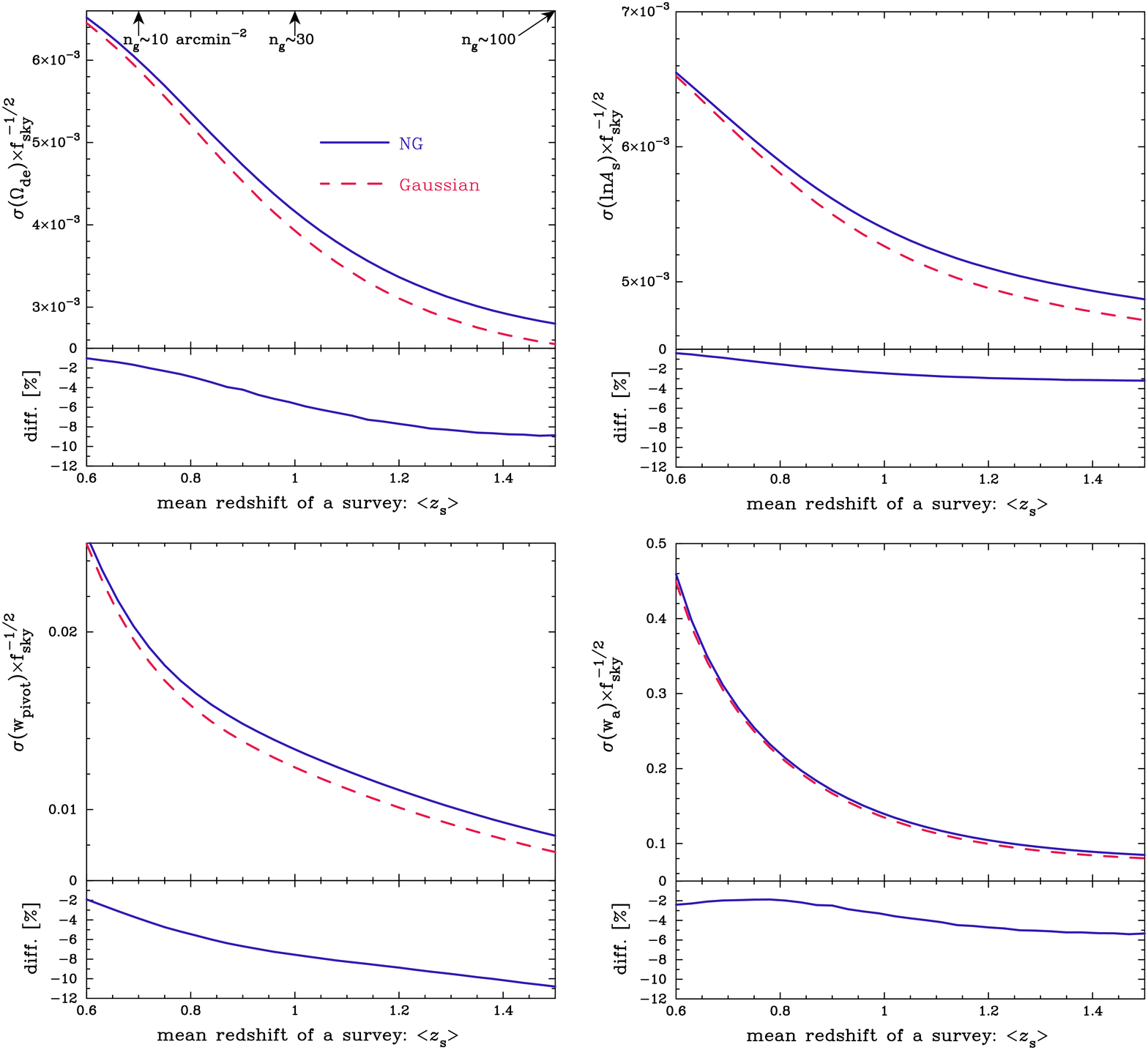}
  \end{center}
\caption{The marginalized $68\%$ errors for 
the parameters $\Omega_{\rm
 de}$, $\ln A_s$ $w_{\rm pivot}$, and $w_a$ as a function of survey
 depth (as in Figure~\ref{fig:sn_zm}). 
These errors are obtained from a 
Fisher matrix analysis combined with the
 Planck priors. 
The parameter
 errors improve with increasing source redshifts, but the error
 degradation due to non-Gaussian errors become larger. 
} \label{fig:errors-zs}
\end{figure}

In Figure~\ref{fig:errors-zs} we study how the impact of the
non-Gaussian contribution on the marginalized error of four parameters
($\Omega_{\rm de}$, $\ln A_s$, $w_{\rm pivot}$, $w_a$) change with
survey depth. It should be noted that, although the parameter errors are
estimated for our fiducial survey area ($2000$ deg$^2$) combined with
the Planck priors, the error shown in the $y$-axis is multiplied by
$f_{\rm sky}^{1/2}\simeq 0.22$ in order to make it easier to reinterpret
the results shown here for any survey area and depth. 
The errors roughly scale with
$f_{\rm sky}^{-1/2}$ even in the presence of the beat-coupling term in
the covariance which does not have this dependence. Although the marginalized
errors for these parameters are improved for higher redshift surveys,
the impact of the non-Gaussian errors are more significant, 
as expected from Figure~\ref{fig:sn_zm}. More encouragingly, 
the
error degradation is generally small, less than $\sim 10\%$, for survey
depth up to $z_m\simeq 1.5$ we have considered here. 

\subsection{Degradations in the Presence of Systematic Errors}
\begin{figure}
  \begin{center}
    \leavevmode\epsfxsize=17.cm \epsfbox{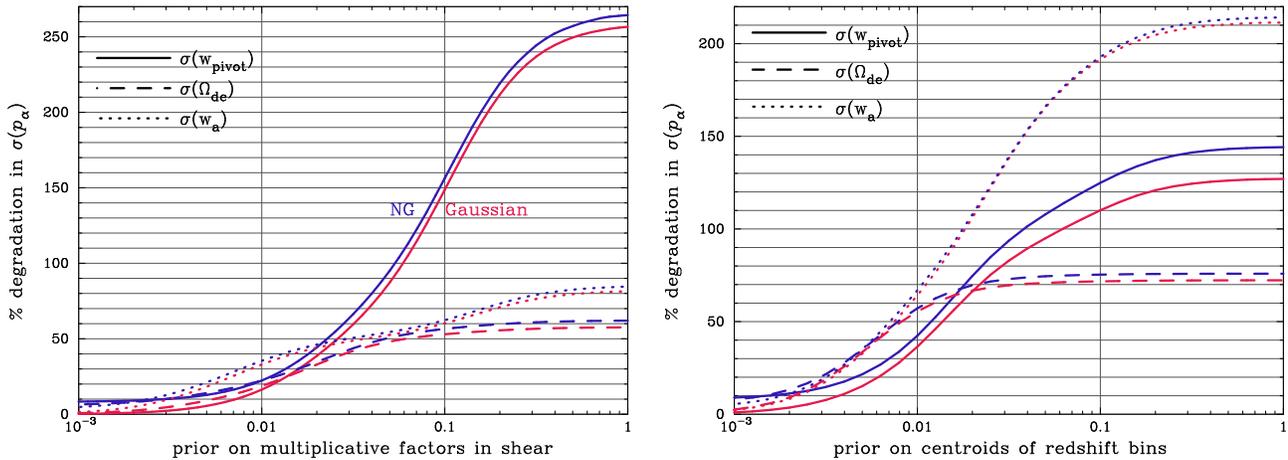}
  \end{center}
\caption{Degradation in the marginalized errors of dark energy
 parameters $\Omega_{\rm de}$, $w_{\rm pivot}$ and $w_a$ as a function
 of the priors on the multiplicative shear errors ({\em Left panel}) or
 the centroids of  redshift bins ({\em Right}). Predictions for 
 both Gaussian and non-Gaussian covariances are shown. 
 Fiducial survey parameters
 and four redshift bins are as in Table~\ref{tab:paras}. For the
 non-Gaussian covariance case, the degradation is relative to the
 Gaussian error results without the systematic errors, so it
 includes the degradation caused by
 non-Gaussian errors  as well as systematic errors. 
 } \label{fig:sys}
\end{figure}

In reality there are various sources of systematic errors that affect weak
lensing measurements.  As discussed in several recent studies (Huterer \&
Takada 2005; Huterer et al. 2006; Ma et al. 2005; Amara \& Refregier
2007; Bernstein 2008), stringent control of those systematic errors is
required in order to retain the ability of a given weak lensing survey
for constraining cosmology. As we have shown, non-Gaussian
correlations between the lensing spectra at different multipoles and
redshift are important, and themselves depend on the underlying cosmological
parameters. On the other hands, the systematic errors generally have
a different dependence on multipole and redshift. Therefore including the
non-Gaussian covariances into the analysis may help discriminate
cosmological signals from systematic errors. 
This is the issue we hope to address in this subsection.

To do this we model the systematic errors following the
method in Huterer et al. (2006): we introduce parameters to describe the
systematic errors and include them as nuisance parameters into the Fisher
analysis. In this paper we consider redshift and multiplicative
shear errors -- the parametrization we use is not generic, but
is expected to account for the salient effects. 

A multiplicative ``calibration bias'' in measuring shear is one of
typical errors that  arise in measuring weak lensing signals
(e.g. Massey et al. 2007). The
general multiplicative error acts on a galaxy image at some redshift
$z_s$ and angular direction $\bm{\theta}$ as $\gamma(z_s,\bm{\theta})
\rightarrow
\gamma(z_s,\bm{\theta})\left[1+\zeta(z_s,\bm{\theta})\right]$, where
$\zeta$ is the multiplicative error (bias) in shear. 
Therefore the lensing spectrum
estimated from galaxies in the $i$- and $j$-th tomographic
redshift bins, $\hat{P}_{(ij)}$, can be biased from the true spectrum
$P_{(ij)}$ as
\begin{equation}
\hat{P}_{(ij)}(l)=P_{(ij)}(l)\left[1+\zeta_i+\zeta_j\right]. 
\end{equation}
Here for simplicity we have assumed that the multiplicative errors
$\zeta$ are not correlated with the cosmological lensing signals, and
are dependent only in the tomographic redshift bins, not on
multipoles, after averaging. We therefore introduce $N_s$
nuisance parameters for the multiplicative shear errors. 

The second systematic error we consider arises from photometric
redshift (photo-$z$) errors. 
The uncertainties in photo-$z$
estimates -- the scatter, bias and fraction of outliers -- may  
significantly degrade the cosmological information of lensing surveys. 
Statistical errors in photo-$z$'s do not by themselves
cause problems for lensing tomography because the tomographic
redshift bins are derived 
from an enormous number of redshifts. Rather it is the mean bias in redshift
bins that leads to systematic errors in lensing. Therefore we introduce
the bias in the centroid of each tomographic redshift bin as a nuisance
parameter. 
We thus introduce $N_s$ nuisance parameters for
lensing tomography with $N_s$ redshift bins. To compute the Fisher
derivatives for these parameters, we vary each centroid by some small
amount $\delta z$ as 
\begin{equation}
\skaco{z_s}_{\mbox{\small $i$-th
redshift bin}}\rightarrow \skaco{z_s}_{\mbox{\small $i$-th redshift
bin}}+\delta z_s
\end{equation}
and then compute the lensing power spectra for the shifted redshift
distribution.

We thus have a total of $9+N_s$ parameters for the two cases, 
multiplicative shear errors and photo-$z$'s errors, and we marginalize
over the $N_s$ nuisance parameters in each case by adding the Gaussian
priors.

Figure~\ref{fig:sys} shows the degradation in the marginalized errors of
$\Omega_{\rm de}$, $w_{\rm pivot}$ and $w_a$ as a function of the priors
on the multiplicative shear errors (left panel) and the redshift bin
centroids (right). We compare the results obtained when the non-Gaussian
errors are included or ignored. Here ``degradation'' means
that the marginalized error for a given parameter is compared to the
error for the Gaussian case without systematic errors:  so it 
shows the effects of both non-Gaussian errors and systematic errors. 
It is evident from Figure~\ref{fig:sys} that in the presence of
these systematic errors, non-Gaussian errors cause very little 
degradation. For the multiplicative shear error, with 
the prior $\sigma(\zeta)=0.01$ chosen to keep the total degradation below
50\%, the differences between the results with and without non-Gaussian
errors 
are about $3$,
$5$ and $2\%$ for $\Omega_{\rm de0}$, $w_{\rm pivot}$ and $w_a$,
respectively.
For the photo-$z$ error, with the prior $\sigma(\delta\!z_s)=0.01$,
those are about $1$, $4$ and $2\%$. 
These numbers can be compared with the results in Table~\ref{tab:paras}
-- 
the degradations
in these dark energy parameters are slightly less significant in the
presence of the systematic errors.
Therefore, including the cosmological
non-Gaussian errors in the covariance 
slightly mitigates requirements on the control of systematic errors 

\section{Covariance of Real-Space Shear Correlations}
\label{sec:2ptcov}

\begin{figure}
  \begin{center}
    \leavevmode\epsfxsize=17.4cm \epsfbox{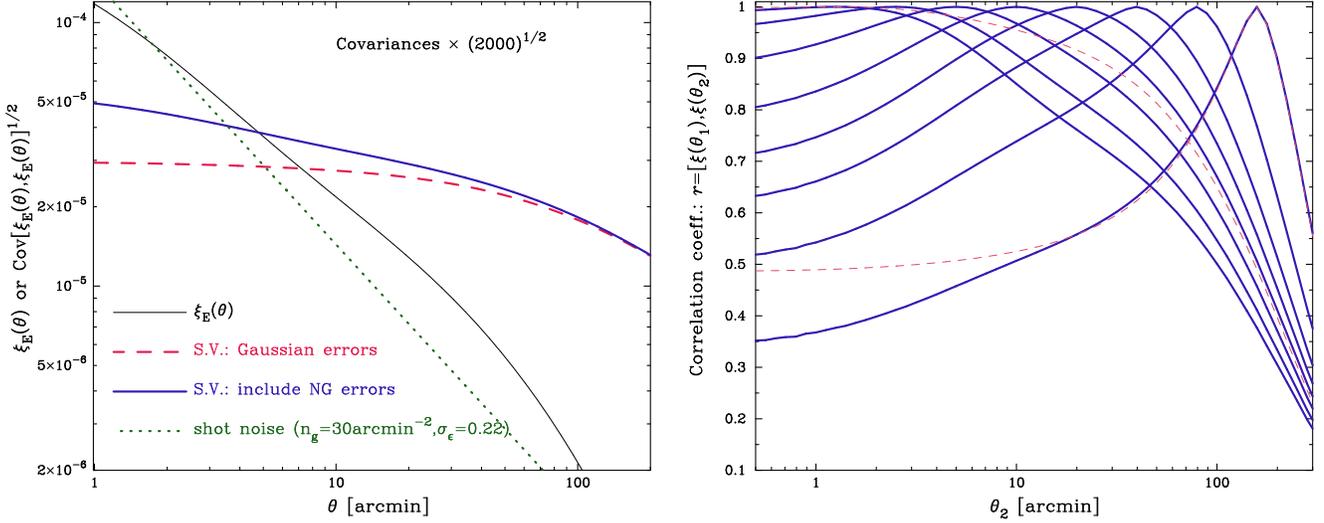}
  \end{center}
\caption{{\em Left panel}: The diagonal components of the covariance
matrix for the shear two-point correlation function as a function of
separation angles. 
The dashed and
bold-solid curves are obtained assuming the
Gaussian  and non-Gaussian errors,
respectively. These can be compared with the two-point correlation
amplitude $\xi_E$. The dotted curve is the shot noise contribution due
to the intrinsic ellipticities. Survey parameters are as in
Figure~\ref{fig:ps}, however, the covariances plotted here are $[{\rm
Cov}\times 2000]^{1/2}$, effectively corresponding to the case for
survey area $1$ deg$^2$, for illustrative purpose. 
{\em Right panel}: The correlation
coefficients $r(\theta_1,\theta_2)$ for the two-point correlation function
defined as in Eqn.~(\ref{eqn:rpij}). 
Note that the assumed $\theta_1$ for each curve is given by the value
of $\theta_2$ at which $r=1$. Shot noise is ignored in
this plot. The two dashed curves show the results when
for the Gaussian error case,
for the smallest and largest angles, $\theta_1\simeq 1.2$ and $160$
arcminutes, respectively. } \label{fig:2ptcov}
\end{figure}
The real-space correlation function of cosmic shear is another
convenient statistic often used in the literature. It has been used in
lensing survey measurements (e.g., Fu et al. 2008), as it does not
require corrections for survey geometry and masking effects.  In this
section we study the covariances of the shear correlation functions. 

The shear $E$-mode correlation function, $\xi_E(\theta)$, is defined in
terms of the lensing power spectrum as
\begin{equation}
\xi_E(\theta)=\int\!\frac{ldl}{2\pi}P(l)J_0(l\theta),
\label{eqn:xie}
\end{equation}
where $J_0(x)$ is the zero-th order Bessel function.  Joachimi et al
(2008) recently developed a useful formula that allows one to compute the
covariances of the real-space correlations in terms of the power
spectrum covariance. Extending this method to the case including the
non-Gaussian error contributions, the covariance of $\xi_E(\theta)$ can
be expressed as
\begin{equation}
{\rm Cov}[\xi_E(\theta),\xi_E(\theta')]\simeq \frac{1}{\pi\Omega_{\rm s}}
\int_0^\infty\!ldl~J_0(l\theta)J_0(l\theta')
P(l)^2
+\frac{1}{4\pi^2\Omega_{\rm s}}
\int_0^\infty\!ldl\int_0^\infty\!l'dl' J_0(l\theta)J_0(l'\theta')
\bar{T}(l,l'), 
\label{eqn:cov2pt}
\end{equation}
where $\bar{T}$ is the angle averaged lensing trispectrum (see the
second term on the r.h.s. of Eqn.[\ref{eqn:pscov}]). 
Note that the beat-coupling effect can be included, e.g. if
Eqn.~(\ref{eqn:pttrisp_bc}) is used to model the perturbation theory
contribution to the trispectrum above.
We consider
no tomography case and ignored shot noise contamination here. 
The first term on the r.h.s. is the Gaussian error,
while the second term gives the non-Gaussian error contribution.  There
are notable differences between the covariances of the power spectrum
and the real-space correlation. Even for a pure Gaussian field, the
first term is non-vanishing for the off-diagonal components of the
covariance when $\theta\ne \theta'$: i.e., the correlation functions of
different angles are always correlated with each other.  
Also note that the Gaussian covariance does not depend
on the bin width of angles. The non-Gaussian errors increase both the
diagonal and off-diagonal terms of the covariance.

Thus the cross-correlations between different angles need to be properly
taken into account when estimating cosmological parameters from the
measured correlation functions. The cosmological interpretation of the
measured correlation functions from the CFHT survey (Fu et al. 2008)
was made using covariances calibrated with ray-tracing
simulations (Semboloni et al. 2007).  Simulation based methods are
indeed needed to study the covariances; however, accurate
calibration of the covariances requires a sufficient number of the
realizations (ideally $\simgt 100$ realizations to attain $\%$ level
accuracy at each bin). In addition the simulations are usually done
assuming some representative cosmological models, such as the concordance
$\Lambda$CDM model. Therefore it is computationally expensive to explore
covariances in parameter space.  Having an analytic method
to compute the covariance is useful and complementary to such simulation
based methods (also see Schneider et al. 2002).

The left panel of Figure~\ref{fig:2ptcov} shows our model predictions
for the diagonal components of the covariance matrix of the shear
correlation function assuming survey parameters as in
Figure~\ref{fig:ps}. For illustrative purpose, the covariance is scaled
as $[{\rm Cov}\times 2000]^{1/2}$, to roughly correspond to 
a survey with area $1$ deg$^2$. The covariance amplitude then
becomes similar to the shear correlation function (thin solid curve).
The dashed and bold-solid curves are the results obtained assuming
Gaussian errors (the first term in Eqn.~[\ref{eqn:cov2pt}]) and 
also including non-Gaussian errors, respectively. 
Note that we ignored the beat-coupling contribution. 
Non-Gaussian errors become more important as one goes to
smaller angular scales, $\theta\simlt 50'$; at the smallest scales
$\theta\sim 1'$ currently probed, non-Gaussian errors increase the
sample variance by a factor of 1.7. This is smaller than the factor of
4 effect found in Semboloni et al. (2007). The difference may be
ascribed partly to the difference in the assumed $\sigma_8$ ($0.8$
vs. $1$), leading to $\sim 20\%$ in ${\rm Cov}^{1/2}$ (see 
Figure~\ref{fig:sn_sig8}).  In addition, it appears that 
they may have underestimated the Gaussian covariance contribution,
which is estimated by integrating the two-point correlations measured
from simulations over a finite range of separation angles. Resolving
this difference in detail is beyond the scope of this paper, but a
detailed study based on simulations would be worth pursuing.  The
dotted curve is the shot noise contamination computed as
\begin{eqnarray}
{\rm Cov}^{\mbox{\scriptsize
shot-noise}}[\xi_E(\theta_i),\xi_E(\theta_j)]
&=&\frac{2\sigma_\epsilon^2}{N_{\rm pair}} \nonumber\\
&\simeq& 2.1\times 10^{-8}\left(\frac{\theta}{1'}\right)^{-2}
\left(\frac{\bar{n}_g}{30 {\rm arcmin}^{-2}}\right)^{-2}
\left(\frac{\Omega_{\rm s}}{1{\rm deg}^2}
\right)^{-1}\left(\frac{\Delta\ln\theta}{0.46}\right)^{-1}
\left(\frac{\sigma_\epsilon}{0.22}\right)^2\delta_{ij}^K,
\label{eqn:cov2pt_sn}
\end{eqnarray}
where $N_{\rm pair}$ is the total number of galaxy pairs, available
from a given survey,  that are
separated by separation angle $\theta$ within bin width $\Delta
\theta$. It can be approximately expressed in terms of
survey parameters as $N_{\rm pair}\simeq
\pi\theta^2\Delta\ln\theta\times \bar{n}_g^2 \Omega_{\rm s}$. The
Kronecker delta function $\delta^K_{ij}$ is needed as the shot noise
contributes only to the diagonal terms of the covariance. 
A factor of $2$ in the numerator of Eqn.~(\ref{eqn:cov2pt_sn}) accounts 
for the fact
that our definition of $\sigma_{\epsilon}$ is for the rms intrinsic
ellipticities per component while the shear $E$-mode correlation arises
from the sum of the two shear components.

The right panel of Figure~\ref{fig:2ptcov} shows the correlation
coefficients for the two-point function, 
defined as in Eqn.~(\ref{eqn:rpij}). Note that the shot noise
contamination, which only contributes to the diagonal terms of the
covariance matrix, is ignored in this plot, and therefore the
off-diagonal correlation coefficients are relatively amplified compared
to the case including the shot noise. Compared to Figure~\ref{fig:rpij}, there
are significant correlations between the two-point correlations of
different separation angles. Also each curve is asymmetric around
the peak, reflecting stronger correlations at smaller scales. 
For comparison, the dashed curves show the results for the Gaussian
error case for two cases of the smallest and largest $\theta_1$'s.
Perhaps surprisingly there are even greater cross-correlations 
than the non-Gaussian error cases, implying the
non-Gaussian errors preferentially contribute to the diagonal
components. 

The results above imply that, if the real-space correlation functions
are used to extract cosmological information,  inclusion of 
cross-correlations between different angles in the analysis is
critically important (e.g. Schneider et al. 2002). 
For example, we will use the
covariance predictions developed in this paper in order
to estimate an upper bound on neutrino masses from the
CFHT lensing data, and have indeed 
found a significant effect of the covariances
on the cosmological parameter constraints (Ichiki, Takada \& Takahashi
2008). 

\section{Discussion}
\label{sec:discuss}

In this paper we have studied how non-Gaussian errors due to nonlinear
clustering affect measurements of weak lensing power spectra. 
We have also estimated the degradation in the accuracies of 
cosmological parameters inferred from 
future lensing surveys. We pay particular attention to a non-Gaussian
contribution that has not been included in the lensing literature so
far, the ``beat coupling'' contribution first studied by 
 Rimes \& Hamilton (2005)
in 3-dimensional simulations. This term vanishes in the infinite area
limit, but for finite survey areas it can dominate the non-Gaussian
contribution 
over some range of angular scales. 
We use two analytical models to compute it. Both 
are based on the halo model and must be tested with
N-body simulations, but based on preliminary comparisons we
expect the simulation results to converge somewhere between the two
models (Figure~\ref{fig:sn_angps} also supports this empirically). 
Our findings can be summarized as follows. 
\begin{itemize}
\item Non-Gaussian covariances significantly increase statistical
      errors on lensing spectra and 
      cause correlations between different multipole bins (see
      Figures~\ref{fig:ps} and \ref{fig:rpij}). 
\item The cumulative signal-to-noise ($S/N$) ratio,
      integrated up to a multipoles $\simgt 1000$, is
      degraded by up to a factor of 2 (or 5 for the worst case scenario)
      compared to the Gaussian case (Figure~\ref{fig:sn_lmax}).
      The degradation is smaller for shallower surveys and when
      including tomographic redshift information (Figure~\ref{fig:sn_zm}). 
\item For parameter estimations from the lensing spectra, 
      non-Gaussian errors enlarge the volume of the error ellipsoid in
      parameter space by a factor that is comparable to
the $S/N$ degradation. 
\item However, since lensing impacts multiple parameters
      the degradation in the marginalized error for individual
      parameters is much smaller. 
      E.g., for an 
      8 parameter Fisher analysis the accuracies of dark energy
      parameters are degraded by less than $10\%\simeq 2^{1/8}-1$ (or
      $20\%$) as shown in  Table~\ref{tab:paras} and Figures~\ref{fig:chi2}, 
      \ref{fig:chi2_nlbc} and \ref{fig:errors-zs}.
\end{itemize}

We have also included two kinds of systematic errors in our analysis:
shear calibration and photo-$z$ bias. Since systematic errors in weak
lensing generally have a different dependence on multipoles and
redshifts, including the non-Gaussian errors in the analysis slightly
mitigates requirements on the control of systematic errors
(Figure~\ref{fig:sys}).
Our model for the power spectrum covariance also allows us to compute the
covariance of shear correlation functions, 
currently used in the analysis of actual measurements. 
This is presented in Section 5. 

The results derived from our analytical method can be used
in making realistic forecasts of the ability of future 
lensing surveys to constrain cosmological parameters. A simulation
based study is needed to test our predictions and choose the detailed
model (Sato et al, in preparation; Pielorz 2008). The effects of
survey geometry and masking can also be estimated from such simulations.
However, numerical estimation of the covariances is 
computationally expense: more than 1000 realizations of ray-tracing
simulations are ideally needed to estimate the covariances at accuracies
better than 10\% (e.g., see Takahashi et al, in preparation).  It
would be even more expensive to study the dependence of the covariances on
cosmological and survey parameters. Therefore, the analytical method
developed here is useful and complementary to such simulation based
studies. For example our method allows one to extend the
covariances for arbitrary cosmological and survey parameters.

The fact that lensing is significantly affected by 
non-Gaussianity implies that the
power spectrum does not carry all the information contained in
lensing fields. Higher-order correlations can be useful
in extracting additional cosmological information.  For example, combining
two- and three-point correlation functions can 
significantly improve  cosmological parameter accuracies (e.g.,
Takada \& Jain 2004). However, one
has to properly take into account covariances of higher-order
moments by including non-Gaussian errors. This requires going up to 
6-point correlations for the bispectrum. We have
found that parameter constraints from 
bisepctrum tomography are degraded more by non-Gaussian covariances
than for the power spectrum. Combining the power spectrum
and bispectrum is still powerful in breaking parameter degeneracies
(Takada \& Jain, in preparation).

Finally, we comment on the possibility of 
combining {\it all} observables available from multicolor imaging
surveys. Besides the lensing 
power spectra, there are various observables that probe
large-scale structure: counting statistics of galaxy clusters,
baryon acoustic oscillations, the full galaxy angular correlation
functions, galaxy-lensing cross-correlation, and so
on. Even though these probe the same cosmic mass density field,
combining these observables can improve accuracies on
cosmological parameter (for example, see Takada \& Bridle
2007). Several open questions remain on strategies for combining
observables
in the presence of systematic
errors.
To address such issue quantitatively, all  covariances between the
observables used have to be correctly taken into account.  Such a
study is challenging, but will be needed to exploit the full
potential of future surveys to constrain the nature of
dark energy or possible modifications of gravity.

\bigskip

We are very grateful to I.~Kayo and R.~Scoccimarro for invaluable
discussions and comments. We also thank G.~Bernstein, S.~Bridle,
D.~Dolney, T.~Hamana, K.~Ichiki, M.~Jarvis, J.~Lee, M.~Kilbinger,
M.~Sato, E.~Sefusatti, E.~Semboloni, P.~Schneider and N.~Yoshida for
useful discussions. We also thank the Lorentz Center at University of
Leiden in Netherlands and IUCAA for their warm hospitality while this
work was initiated.  MT also thanks the members at Astronomy Department
of University of Pennsylvania for their warm hospitality.  This work was
in part supported by World Premier International Research Center
Initiative (WPI Initiative), MEXT, Japan, by Grand-in-Aid for Scientific
Research on Priority Area No. 467 ``Probing Dark Energy through an
Extremely Wide and Deep Survey with Subaru Telescope'' and on young
researchers (Nos. 17740129 and 20740119) as well as by Grand-in-Aid for
the 21st Century Center of Excellence program ``Exploring New Science by
Bridging Particle-Matter Hierarchy'' at Tohoku University. BJ is
supported in part by NSF grant AST-0607667. 

\appendix

\section{Derivation of the beat-coupling contribution to 
the lensing covariance}
\label{app:bc}

In this section we derive the beat-coupling contribution to the power
spectrum covariance in more detail (also see Hamilton et al. 2006 and
Sefusatti et al. 2006). 

Let us begin by recalling the definition of the 3D mass trispectrum
in terms of the mass fluctuation field $\delta_m(\bm{k})$:
\begin{equation}
\skaco{\delta_m\!(\bm{k}_1)\delta_m\!(\bm{k}_2)\delta_m\!(\bm{k}_3)
\delta_m\!(\bm{k}_4)
}\equiv
(2\pi)^3\delta_D(\bm{k}_{1234})T_\delta\!(\bm{k}_1,\bm{k}_2,\bm{k}_3,
\bm{k}_4), 
\label{eqn:mtrisp}
\end{equation}
where we have introduced notation such as $\bm{k}_{1234}\equiv
\bm{k}_1+\bm{k}_2+\bm{k}_3+\bm{k}_4$, and the Dirac delta function
imposes the condition that the four wavevectors
$\bm{k}_1,\cdots,\bm{k}_4$ form a closed 4-point configuration in
Fourier space. According to perturbation theory (e.g. see Bernardeau
et al. 2002 for a thorough review), the mass trispectrum can be
expressed in terms of the linear mass power spectrum as
\begin{eqnarray}
T_\delta^{\rm PT}(\bm{k}_1,\bm{k}_2,\bm{k}_3,\bm{k}_4) &=&4\left[
F_2(\bm{k}_{12},-\bm{k}_1)F_2(\bm{k}_{12},\bm{k}_3)P^L_\delta\!(k_{12})P^L_\delta\!(k_1)P^L_\delta\!(k_3)
+F_2(\bm{k}_{12},-\bm{k}_1)F_2(\bm{k}_{12},\bm{k}_4)P^L_\delta\!(k_{12})P^L_\delta\!(k_1)P^L_\delta\!(k_4)
\right.  \nonumber\\
&&\hspace{-0em}+F_2(\bm{k}_{12},-\bm{k}_2)F_2(\bm{k}_{12},\bm{k}_3)P^L_\delta\!(k_{12})P^L_\delta\!(k_2)P^L_\delta\!(k_3)
+F_2(\bm{k}_{12},-\bm{k}_2)F_2(\bm{k}_{12},\bm{k}_4)P^L_\delta\!(k_{12})P^L_\delta\!(k_2)P^L_\delta\!(k_4)
\nonumber\\ &&\hspace{-0em}
+F_2(\bm{k}_{13},-\bm{k}_1)F_2(\bm{k}_{13},\bm{k}_2)P^L_\delta\!(k_{13})P^L_\delta\!(k_1)P^L_\delta\!(k_2)
+F_2(\bm{k}_{13},-\bm{k}_1)F_2(\bm{k}_{13},\bm{k}_4)P^L_\delta\!(k_{13})P^L_\delta\!(k_1)P^L_\delta\!(k_4)
\nonumber\\ &&\hspace{-0em}
+F_2(\bm{k}_{13},-\bm{k}_3)F_2(\bm{k}_{13},\bm{k}_2)P^L_\delta\!(k_{13})P^L_\delta\!(k_3)P^L_\delta\!(k_2)
+F_2(\bm{k}_{13},-\bm{k}_3)F_2(\bm{k}_{13},\bm{k}_4)P^L_\delta\!(k_{13})P^L_\delta\!(k_3)P^L_\delta\!(k_4)
\nonumber\\ &&\hspace{-0em}
+F_2(\bm{k}_{14},-\bm{k}_1)F_2(\bm{k}_{14},\bm{k}_2)P^L_\delta\!(k_{14})P^L_\delta\!(k_1)P^L_\delta\!(k_2)
+F_2(\bm{k}_{14},-\bm{k}_1)F_2(\bm{k}_{14},\bm{k}_3)P^L_\delta\!(k_{14})P^L_\delta\!(k_1)P^L_\delta\!(k_3)
\nonumber\\ &&\hspace{-0em} \left.
+F_2(\bm{k}_{14},-\bm{k}_4)F_2(\bm{k}_{14},\bm{k}_2)P^L_\delta\!(k_{14})P^L_\delta\!(k_4)P^L_\delta\!(k_2)
+F_2(\bm{k}_{14},-\bm{k}_4)F_2(\bm{k}_{14},\bm{k}_3)P^L_\delta\!(k_{14})P^L_\delta\!(k_4)P^L_\delta\!(k_3)
\right] \nonumber\\ &&\hspace{0em} +6\left[
F_3(\bm{k}_1,\bm{k}_2,\bm{k}_3)P^L_\delta\!(k_1)P^L_\delta\!(k_2)P^L_\delta\!(k_3)
+F_3(\bm{k}_1,\bm{k}_2,\bm{k}_4)P^L_\delta\!(k_1)P^L_\delta\!(k_2)P^L_\delta\!(k_4)
+\mbox{3 perms.}\right],
\label{eqn:pttrisp}
\end{eqnarray}
where the kernels $F_2$ and $F_3$ are given by
\begin{eqnarray}
F_2(\bm{k}_1,\bm{k}_2)&=&\frac{5}{7}+\frac{1}{2}\left(
\frac{k_1}{k_2}+\frac{k_2}{k_1}
\right)\frac{\bm{k}_1\cdot\bm{k}_2}{k_1k_2}
+\frac{2}{7}\frac{(\bm{k}_1\cdot\bm{k}_2)^2}{k_1^2k_2^2}
\label{eqn:f2}
\\
F_3(\bm{k}_1,\bm{k}_2,\bm{k}_3)&=&
\frac{1}{54}\left[
7\frac{\bm{k}_{123}\cdot\bm{k}_1}{k_1^2}
F_2(\bm{k}_2,\bm{k}_3)+
\left\{
7\frac{\bm{k}_{123}\cdot\bm{k}_{23}}{k^2_{23}}+2\frac{k_{123}^2(\bm{k}_{23}\cdot\bm{k}_1)}{{k_{23}^2k_1^2}}
\right\}G_2(\bm{k}_2,\bm{k}_3)+\mbox{2 perms.}
\right]
\end{eqnarray}
with another kernel $G_2$ defined as 
\begin{equation}
G_2(\bm{k}_1,\bm{k}_2)\equiv \frac{3}{7}+\frac{1}{2}\left(
\frac{k_1}{k_2}+\frac{k_2}{k_1}
\right)\frac{\bm{k}_1\cdot\bm{k}_2}{k_1k_2}
+\frac{4}{7}
\frac{(\bm{k}_1\cdot\bm{k}_2)^2}{k_1^2k_2^2}.
\end{equation}
Note that, due to the condition
$\bm{k}_1+\bm{k}_2+\bm{k}_3+\bm{k}_4=\bm{0}$, there are many other ways
to express the mass trispectrum (\ref{eqn:pttrisp}).
For convenient purpose of our following
discussion, we express the mass trispectrum such that all the term
has arguments of $\bm{k}_{1a}$ where $a=2, 3$ or $4$.

The covariance of the mass power spectra at two wevenumbers $k$ and $k'$
arises from the 4-point correlations of $\delta_m$ with specific
configurations in Fourier space: $\skaco{P_\delta\!(k)P_\delta\!(k')} =
\skaco{\delta_m(\bm{k})
\delta_m(-\bm{k})\delta_m(\bm{k}')\delta_m(-\bm{k}') }$, because of
$P_\delta\!(k)=\skaco{\delta_m(\bm{k})\delta_m(-\bm{k}) }$ and so
on. Note that, even for the diagonal terms of the covariance where
$k=k'$, the two vectors $\bm{k}$ and $\bm{k}'$ generally differ in
direction.  As discussed in \S~\ref{sec:bc}, for any survey of a finite
sky or volume coverage, we cannot measure Fourier modes to a better
accuracy than the fundamental model of the survey, say
$\varepsilon_k=2\pi/D$, where $D$ is a linear scale of the survey
region: two modes that differ by $\bm{\varepsilon}_k$ cannot be in
practice distinguished due to the limited resolution. Taking into
account this uncertainty, the PT trispectrum contribution to the
non-Gaussian part of the mass power spectrum covariance arises from the
following mass trispectrum (also see Hamilton et al. 2006):
\begin{equation}
{\rm Cov}^{{\rm NG, PT}}[P_\delta(k),P_\delta(k')]
\longleftarrow T^{\rm PT}_\delta(\bm{k}+\bm{\varepsilon}'_k,
-\bm{k}+\bm{\varepsilon}_k^{\prime\prime},
\bm{k}'+\bm{\varepsilon}^{\prime\prime\prime}_k,
-\bm{k}'+\bm{\varepsilon}_k^{\prime\prime\prime\prime}
),\label{eqn:bc_pt}
\end{equation}
where the vectors $\bm{\varepsilon}_k$ with prime superscripts denote
the fundamental modes of amplitude $2\pi/D$.

Assuming the case that the modes of interest are much greater than the
fundamental modes, e.g. $k,k'\gg \varepsilon_k'$, substituting
Eqn.~(\ref{eqn:bc_pt}) into Eqn.~(\ref{eqn:pttrisp}) yields
\begin{eqnarray}
 T^{\rm PT}_\delta\!(\bm{k}+\bm{\varepsilon}'_k,
-\bm{k}+\bm{\varepsilon}_k^{\prime\prime},
\bm{k}'+\bm{\varepsilon}^{\prime\prime\prime}_k,
-\bm{k}'+\bm{\varepsilon}_k^{\prime\prime\prime\prime})&\approx&
T^{\rm PT}_\delta\!(\bm{k},-\bm{k},\bm{k}',-\bm{k}')\nonumber\\
&&\hspace{-20em}+4P^L_\delta\!(\varepsilon_k)P^L_\delta\!(k)
P^L_\delta\!(k')\left[
F_2(\bm{\varepsilon}_k,-\bm{k})F_2(\bm{\varepsilon}_k,\bm{k}')
+F_2(\bm{\varepsilon}_k,-\bm{k})F_2(\bm{\varepsilon}_k,
-\bm{k}')
+F_2(\bm{\varepsilon}_k,\bm{k})F_2(\bm{\varepsilon}_k,\bm{k}')
+F_2(\bm{\varepsilon}_k,\bm{k})F_2(\bm{\varepsilon}_k,-\bm{k}')
\right], 
\label{eqn:mtrisp_bc}
\end{eqnarray}
where $\bm{\varepsilon}_k\equiv \bm{\varepsilon}_k^\prime+
\bm{\varepsilon}_k^{\prime\prime}$. In the derivation above, we have
used the fact that the first four terms in the square bracket on the
r.h.s. of Eqn.~(\ref{eqn:pttrisp}) can be computed as
$F_2(\bm{k}_{12},-\bm{k}_1)P^L_\delta\!(k_{12}) P^L_\delta\!(k_1)
\approx
F_2(\bm{\varepsilon}^\prime+\bm{\varepsilon}_k^{\prime\prime},-\bm{k})
P^L_\delta\!(|\bm{\varepsilon}_k^\prime+\bm{\varepsilon}_k^{\prime\prime}
|)P^L_\delta(k)$ when $\bm{k}_1=\bm{k}+\bm{\varepsilon}^\prime$ and
$\bm{k}_2=-\bm{k}+ \bm{\varepsilon}^{\prime\prime}$, which are thus
proportional to $P(\varepsilon_k)$ and greater than the other terms
because $F_2\sim O(1)$ and $P^L_{\delta}\!(\varepsilon_k)\gg
P^L_{\delta}\!(k)$ for a CDM spectrum. The other terms are computed as
$F_2(\bm{k}_{13},-\bm{k}_1)P^L_\delta\!(k_{13})P^L_\delta\!(k_1)\approx
F_2(|\bm{k}+\bm{k}'|,-\bm{k})P_\delta^L(|\bm{k}+\bm{k}'|)P_\delta^L\!(k)$
under the assumption $k\gg \varepsilon_k$, and are rewritten as the form
of the standard trispectrum contribution, $T^{\rm PT}_\delta\!(\bm{k},
-\bm{k},\bm{k}',-\bm{k}')$. By integrating the equation above over
angles of wavevectors $\bm{k}$ and $\bm{k}'$, combined with the Limber's
approximation and the lensing projection, the contribution to the lensing
covariance given by Eqn.~(\ref{eqn:lensng_pt}) can be obtained.

Although a rather mathematical derivation of the beat-coupling was
described above, we will in the following make a more intuitive
explanation on the beat-coupling effect within the framework of the halo
model approach. In the halo model picture, the mass density fluctuation
field of a given wavevector $\bm{k}$ is expressed as the sum of the mass
fluctuations in the highly nonlinear regime, confined within one halo,
and the mass fluctuations in the weakly nonlinear regime arising from
the halo distribution, which we here refer as to the 1-halo term,
$\delta_m^{1h}$, and the perturbation theory contribution, $\delta^{\rm
PT}_m$, respectively:
\begin{equation}
\delta_m\!(\bm{k})=\delta_m^{\rm PT}\!(\bm{k})+\delta_m^{1h}\!(\bm{k}). 
\end{equation}
In the presence of the fundamental mode uncertainty the mass
fluctuations in the weakly nonlinear regime contain contributions of
physical correlations between the modes of $\bm{k}$ and
$\bm{\varepsilon}$ arising from nonlinearities of gravitational
clustering. More precisely, based on the perturbation theory, the
density fluctuation field $\delta^{\rm PT}_m$ can be expanded as
\begin{eqnarray}
\delta_m^{\rm PT}\!(\bm{k}+\bm{\varepsilon}_k)
&=& \delta_m^{(1)}\!(\bm{k}+\bm{\varepsilon}_k)
+\delta_m^{(2)}\!(\bm{k}+\bm{\varepsilon}_k)
+\delta_m^{(3)}\!(\bm{k}+\bm{\varepsilon}_k)+\cdots
\nonumber \\
&\approx& \delta_m^{(1)}\!(\bm{k})
+\int\!\frac{d^3\bm{q}}{(2\pi)^3} F_2(\bm{q},\bm{q}-\bm{k}-\bm{\varepsilon}_k)
\delta^{(1)}_m\!(\bm{q})\delta^{(1)}_m\!(\bm{k}+\bm{\varepsilon}_k-\bm{q})
\nonumber\\
&&+\int\!\frac{d^3\bm{q}_1}{(2\pi)^3} \frac{d^3\bm{q}_2}{(2\pi)^3} 
F_3(\bm{q}_1,\bm{q}_2,\bm{k}+\bm{\varepsilon}_k-\bm{q}_1-\bm{q}_2)
\delta^{(1)}_m\!(\bm{q}_1)\delta^{(1)}_m\!(\bm{q}_2)
\delta^{(1)}_m\!(\bm{k}+\bm{\varepsilon}_k-\bm{q}_1-\bm{q}_2)
+\cdots,
\end{eqnarray}
where $\delta^{(1)}_m$, $\delta_m^{(2)}$ and $\delta_m^{(3)}$ are the
linear-, 2nd- and 3rd-order contributions of the mass fluctuations in
the perturbative expansion (e.g., see Jain \& Bertschinger 1994), and in
the second equality on the r.h.s. of the equation above we have used
that the linear-order fluctuations are only slowly varying with
wavenumbers for a CDM model,
$\delta^{(1)}_m\!(\bm{k}+\bm{\varepsilon}_k)\approx
\delta^{(1)}_m\!(\bm{k})$. Thus gravitational instability that doesn't
have any characteristic scale predicts that all the fluctuations of
different scales are coupled to each other. In particular, the 2nd-order
density fluctuations are found to contain contributions arising from the
correlations between the fluctuations of $\bm{k} $ and the fundamental
mode: $F_2(\bm{k},\bm{\varepsilon}_k)\delta^{(1)}_m\!
(\bm{k})\delta^{(1)}_m\!(\bm{\varepsilon}_k)$ when $\bm{q}=\bm{k}$.
Such contributions to the power spectrum covariance indeed arise after
taking the ensemble average such as
$\skaco{\delta^{(1)}_m\!(-\bm{k})\delta^{(1)}\!(\bm{q})
\delta^{(1)}_m\!(\bm{k}+\bm{\varepsilon}_k-\bm{q})} $ in the covariance
calculation.
Thus the beat-coupling contribution is caused by the physical
correlation between the mass fluctuations of large- and small-distance
scales as predicted by the perturbation theory of mass clustering. In
other words, if there are no such correlations, e.g. in the case of the
linear regime, the power spectrum measured from a finite survey is not
influenced by the fundamental mode uncertainty.

On the other hand, the highly nonlinear mass fluctuations are unlikely
to be affected by the fundamental mode uncertainty as follows. Since the
mass fluctuations in the highly nonlinear regime are very likely to lie
inside a halo, the self-gravitating bound object, the mass distribution
within the halo would be sufficiently decoupled from and unaffected by
large-scale fluctuations, for example as in the stable clustering ansatz
where the highly nonlinear fluctuations are assumed to be totally
decoupled from the Hubble flow. Therefore the 1-halo trispectrum
contribution to the power spectrum covariance is unlikely to be
contaminated by the beat-coupling effect. Note that, although the 1-halo
term also depends on the halo mass function in addition to the mass
distribution inside a halo, the mass function can be mapped out from the
{\em linear-order} mass fluctuations to zero-th order approximation, as
done in the Press-Schechter prescription, and therefore would not be
affected by the fundamental mode uncertainty according to the rationale
discussed in the previous paragraph.

\section{Covariances of Angular Mass Power Spectrum: Comparison of our
 model prediction with Lee \& Pen}
\label{app:leepen}
\begin{figure}
  \begin{center}
    \leavevmode\epsfxsize=10.cm \epsfbox{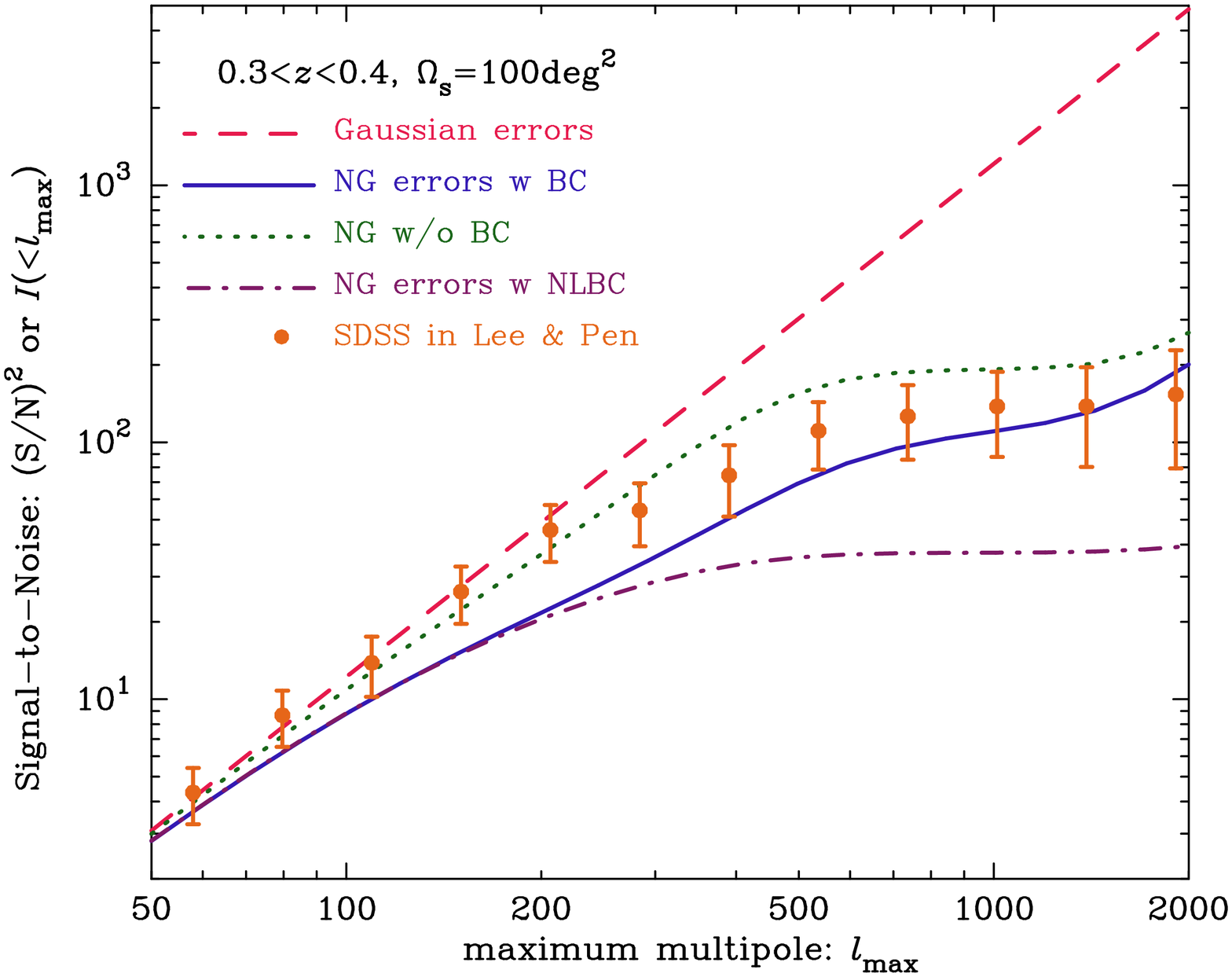}
  \end{center}
\caption{Comparison of our model predictions with the SDSS measurements
of Lee \& Pen (2008), for the cumulative signal-to-noise (or
the information content) of the mass or galaxy angular power spectrum,
respectively. The survey parameters are taken to resemble the SDSS
result. As in Figure~\ref{fig:sn_lmax}, the solid and dotted curves
are the results obtained including the non-Gaussian errors with and
without the beat-coupling effect, while the dashed curve is the result
for the Gaussian error case, which scales as $(S/N)^2\propto l_{\rm
max}^2$. Our model predictions taking account of non-Gaussian
errors are in fairly good agreement with the measurement result,
even though there is uncertainty due to galaxy bias (see text). 
 The dot-dashed curve shows the result assuming the
nonlinear beat-coupling, which is  smaller than the measurement by a
factor of a few.} \label{fig:sn_angps}
\end{figure}
In this appendix, as one check of our model,
we will show the comparison of our predictions
with the SDSS angular galaxy power spectrum recently measured in Lee \&
Pen (2008).

The formulation we have developed in this paper is readily applied to
the angular power spectrum of mass clustering if the redshift weight
function for lensing, $W_{(i)}(\chi)$ (e.g. see Eqn.~[\ref{eqn:ps}]), is
replaced with the selection function of a given galaxy survey as
\begin{equation}
W_{(i)}(\chi)\rightarrow p_g(z)\frac{dz}{d\chi}, 
\end{equation}
where $p_g(z)$ is the selection function: $p_g(z)$ is unity if $z$ is in
the range of redshifts surveyed, otherwise zero. Note that the selection
function should be normalized so as to satisfy $\int\!dz~p_g(z)=1$.

Figure~\ref{fig:sn_angps} shows the cumulative signal-to-noise ratio expected
for measuring angular power spectrum of the mass distribution over
multipole range $2\le l \le l_{\rm max}$ as a function of the maximum
multipole $l_{\rm max}$, assuming a survey with area $\Omega_{\rm
s}=100$ deg$^2$ and redshift coverage $0.3\le z\le 0.4$ that are chosen
to resemble the survey parameters in Lee \& Pen.  Here we didn't include
the shot noise contamination such that our $(S/N)^2$ given by
Eqn.~(\ref{eqn:snps1}) becomes equivalent to the Fisher information
content $I$ studied in Lee \& Pen, where the shot
noise contamination due to discrete
galaxy distribution is subtracted.  It should be
also worth noting that the galaxy bias uncertainty may not so largely
change our results for mass power spectrum, as the bias factor to some
extent cancels out in the $(S/N)^2$ evaluation. For the linear bias
case, the $(S/N)^2$'s for mass and galaxy distribution are exactly
equivalent as $P_g=b^2P_\delta$ and $[{\rm Cov}(P_g)]^{-1}=b^{-4} [{\rm
Cov}(P_\delta)]^{-1} $.

The solid and dotted curves show our model predictions including
non-Gaussian errors, but with and without the beat-coupling
effect, respectively, while the dashed curve show the result for
the Gaussian error, which 
scales as $l_{\rm max}^2$ in the absence of shot
noise. Several interesting points can be found from 
this plot. First, the
beat-coupling effect appears to be more significant for this angular
power spectrum than the lensing spectrum, because the angular power
spectrum has a much narrower redshift coverage, i.e. less line-of-sight
projection. More precisely, since the comoving
angular diameter distance to $z\simeq 0.35$ is $\chi\simeq 1400$Mpc for
our fiducial cosmology, the translinear regime of $k\simeq
[0.1,1]$Mpc$^{-1}$, where the beat-coupling is expected to be 
significant (Rimes \& Hamilton 2005), appears over a wider range of
multipoles $l\simeq k\chi\simeq [140, 1400]$. For the lensing case, the
translinear regime signature is smeared out by the projection over a
wider range of redshifts. Second, very encouragingly, our model
predictions fairly well reproduce the measurement results.
On the other hand, the dot-dashed curve show the
the result obtained assuming the nonlinear beat-coupling effect. This
prediction underestimates the measurement by a factor of 3 on $l_{\rm
max }\simgt 500$, implying an overestimation of the beat-coupling effect
in this prescription. 

\section{Dependence of the lensing $S/N$ on other parameters}
\label{app:sn}

\begin{figure}
  \begin{center}
    \leavevmode\epsfxsize=18.cm \epsfbox{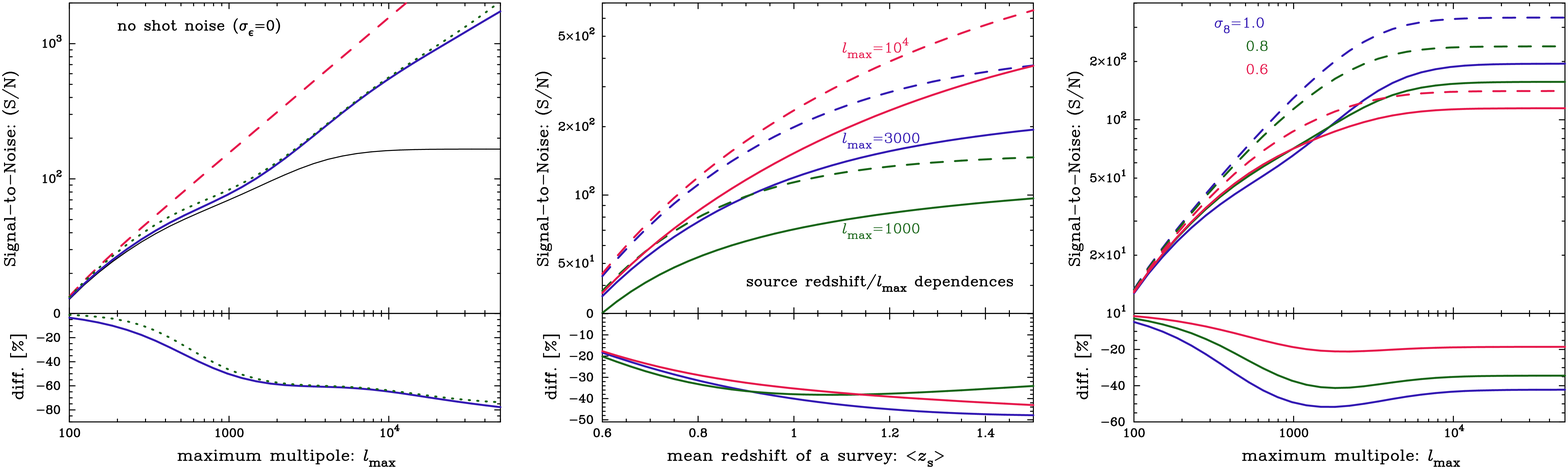}
  \end{center}
\caption{{\em Left panel}:
As in Figure~\ref{fig:sn_lmax}, but
 the shot noise due to the intrinsic ellipticities is ignored. For
 reference, the thin-solid curve shows the result shown by the
 bold-solid curve in Figure~\ref{fig:sn_lmax}. Note that the $y$-axis range in
 the middle and right panels are different from those in the left panel. 
{\em Middle panel}: As in the left panel of
 Fig.~\ref{fig:sn_zm}, but for three different choices of the maximum
 multipole, $l_{\rm max}=1000$, 3000 and $10^4$. The dashed and solid
 curves for each case of $l_{\rm max}$ show the results for Gaussian
 and non-Gaussian errors, respectively.  {\em Right panel}: 
The dependence of the total $S/N$ on
 $\sigma_8$ is shown. The $S/N$ increases with increasing
 $\sigma_8$ for $l_{\rm max}\simgt 2000$, but the non-Gaussian errors
degrade the $S/N$ more strongly, due to stronger nonlinear
clustering. 
}  \label{fig:sn_sig8}
\end{figure}

In this appendix we study how the impact of non-Gaussian errors on 
the $S/N$ for the lensing measurement
depends on shot noise, multipole range, source redshift and
cosmological parameters. 

The shot noise of intrinsic ellipticities
contributes only to the diagonal terms of the
power spectrum covariances. Hence switching off the shot noise terms in
the covariances, corresponding to the case of an infinite number density
of source galaxies, enhances the effect of non-Gaussian errors on the $S/N$,
which is studied in the left panel of Figure~\ref{fig:sn_sig8}.  Note
that, in this case, the $S/N$ value for the Gaussian error case scales
with $l_{\rm max}$ as
$S/N\propto l_{\rm max}$. Compared to Figure~\ref{fig:sn_lmax}, 
the $S/N$ values
are significantly boosted, e.g. by an order of magnitude on $l_{\rm
max}\simgt 10^{4}$ for the Gaussian error case. Comparing the lower
panels of this plot and Figure~\ref{fig:sn_lmax}
also manifests that 
non-Gaussian errors more degrades the $S/N$ when the shot noise is
ignored.  Hence, the shot noise is found to not only significantly
reduce the total $S/N$ at $l_{\rm max}\simgt 1000$, but also mitigate
the impact of the non-Gaussian errors.

The middle pane shows similar results to Figure~\ref{fig:sn_zm}, but
for
different choices of $l_{\rm max}$; $l_{\rm max}=1000$, 3000 and $10^4$,
respectively. The impact of the non-Gaussian errors appears to be very
similar for these $l_{\rm max}$.

The lensing signals as well as the non-Gaussian errors are both
sensitive to strengths of nonlinear mass clustering that is
characterized by $\sigma_8$.  The right panel of Figure~\ref{fig:sn_sig8}
shows how the $S/N$ varies with $\sigma_8$ assumed, but other parameters
being fixed.
Note that our fiducial model has
$\sigma_8\simeq 0.8$. For the Gaussian error cases, the $S/N$ amplitudes
increase with increasing $\sigma_8$ as expected, due to the reduced shot
noise in relative by the enhanced lensing signals. For the
non-Gaussian error cases, however, the $S/N$ amplitudes are more degraded with
increasing $\sigma_8$ due to stronger nonlinearities of mass clustering, 
as explicitly demonstrated in the lower panel. An even more
interesting is, over a range of multipoles $200\simlt l_{\rm max}\simlt
1000$, the total $S/N$ amplitudes decrease with $\sigma_8 $ due to stronger
effects of the non-Gaussian errors. Thus this modestly large dependence
of the non-Gaussian covariances on $\sigma_8$ would need to be realized,
if the covariances are calibrated based on simulations assuming some
$\sigma_8$ that is different from the true cosmology. For example,
Semboloni et al. (2007) studied the non-Gaussian covariances for the
real-space cosmic shear correlation function using simulations with
$\sigma_8=1$, so the non-Gaussian effect may be overestimated if 
the universe has a smaller $\sigma_8$ such as $\sigma_8\approx 0.8$.

\label{lastpage}


\begin{thebibliography}{}

%
\bibitem[Albrecht et al. 2006]{DETF}
Albrecht, A., et al., 2006, astro-ph/0609591

%
\bibitem[Amara \& Refregier 2007]{AmaraRefregier07}
Amara, A., Refregier, A., 2007, arXiv:0710.5171

%
\bibitem[Bartelmann \& Schneider 2001]{BartelmannSchneider}
Bartelmann, M., Schneider, P., 2001, Phys.~Rep., 340,  291


%
\bibitem[Bernardeau et al. 2002]{Bernardeauetal02}
Bernardeau, F., Colombi, S., Gazta\~naga, E., Scoccimarro, R., 
2002, Phys.~Rep., 367, 1 

%
\bibitem[Bernstein08]{Bernstein08}
Bernstein, G.~M., arXiv:0808.3400

%
\bibitem[Bernstein \& Jain 2004]{BernsteinJain04}
Bernstein, G.~M., Jain, B., 2004, ApJ, 600, 17

%
\bibitem[Cooray \& Hu 2001]{CoorayHu01}
Cooray, A., Hu, W., 2001, ApJ, 554, 56

%
\bibitem[Cooray \& Sheth 2002]{CooraySheth02}
Cooray, A., Sheth, R., 2002, Phys.~Rept., 327, 1

%
\bibitem[Dodelson 03]{Dodelson03}
Dodelson, S., 
{\em Modern Cosmology}, Academic Press, San Diego
    (2003).

%
\bibitem[Eisenstein \& Hu 1999]{EisensteinHu99}
Eisenstein, D.~J., Hu, W., 1999, ApJ, 511, 5

%
\bibitem[Eisenstein et al. 1998]{Eisenstein}
Eisenstein, D.~J., Hu, W., Tegmark, M., 1998, ApJ, 518, 2

%
\bibitem[Fosalba, Pan \& Szapudi]{Fosalbaetal05}
Fosalba, P., Pan, J., Szapudi, I., 2005, ApJ, 632, 29

%
\bibitem[Fu et al. 2008]{Fuetal08}
Fu, L., et al., arXiv:0712.0884

%
\bibitem[Hamilton et al. 2006]{Hamiltonetal06}
Hamilton, A.~J.~S., Rimes, C.~D., Scoccimarro, R., 
2006, MNRAS, 371, 1188

%
\bibitem[Heavens 03]{Heavens03}
Heavens, A.~F., 2003, MNRAS, 343, 1327


%
\bibitem[Hoekstra \& Jain 2008]{HoekstraJain08}
Hoekstra, H., Jain, B.,  arXiv:0805.0139

%
\bibitem[Hu 1999]{Hu99}
Hu, W., 1999, ApJ, 522, L21

%
\bibitem[Hu 2000]{Hu00}
Hu, W., 2000, Phys.~Rev.~D, 62, 3007



%
\bibitem[Huterer 2002]{Huterer02}
Huterer, D., 2002, Phys.~Rev.~D., 65, 063001

%
\bibitem[Huterer \& Takada 2005]{HutererTakada05}
Huterer, D., Takada, M., 2005, Astropart.~Phys., 23, 369

%
\bibitem[Huterer et al. 2006]{Hutereretal06}
Huterer, D., Takada, M., Bernstein, G., Jain, B., 2006, MNRAS, 366, 101

%
\bibitem[Ichiki et al. 2008]{Ichikietal08}
Ichiki, K., Takada, M., Takahashi, T.,  arXiv:0810.4921

%
\bibitem[Jain \& Bertschinger 1994]{JainBertschinger94}
Jain, B., Bertschinger, E., 1994, ApJ, 431, 486





%
\bibitem[Joachimi et al. 2008]{Joachimietal08}
Joachimi, B., Schneider, P., Eifler, T., 2008, A\&A, 477, 43


%
\bibitem[Komatsu et al. 2008]{WMAP}
Komatsu, E., et al.,  arXiv:0803.0547

%
\bibitem[Lee \& Pen 2008]{LeePen08}
Lee, J., Pen, U.-L., 2008, ApJ, 686, L1

%
\bibitem[Lewis et al. 2000]{CAMB}
Lewis, A., Challinor, A., Lasenby, A., 2000, ApJ, 538, 473

%
\bibitem[Limber 1954]{Limber54}
Limber, D.~N., 1954, ApJ, 119, L655

%
\bibitem[Ma 2007]{Ma07}
Ma, Z., 2007, ApJ, 665, 887

%
\bibitem[Ma et al. 2005]{Maetal05}
Ma, Z., Hu, W., Huterer, D., 2005, ApJ, 636, 21

%
\bibitem[Ma \& Fry 2000]{MaFry00}
Ma, C.-P., Fry, J.~N., 2000, ApJ, 543, 503

%
\bibitem[Makino, Sasaki \& Suto 1992]{Makinoetal92}
Makino, N., Sasaki, M., Suto, Y., 1992, Phys.~Rev.~D, 46, 585

%
\bibitem[Massey et al. 2007]{Masseyetal07}
Massey, R., et al., 2007, MNRAS, 376, 13 

%
\bibitem[Mellier 1999]{Mellier99}
Mellier, Y., 1999, ARAA, 37, 127

%
\bibitem[Miyazaki et al. 2006]{Miyazaki06}
Miyazaki, S., et al. 2006 SPIE 6269, 9

%
\bibitem[Neyrinck et al 2006]{Neyrincketal06}
Neyrinck, M.~C., Szapudi, I., Rimes, C.~D., 2006, MNRAS, 370, L66

%
\bibitem[Peacock \& Smith 2000]{PeacockSmith00}
Peacock, J.~A., Smith, R.~E., 2000, MNRAS, 318, 1144



%
\bibitem[Perlmutter et al. 1999]{Perl}
Perlmutter, S., et al., 1999, ApJ, 517, 565


\bibitem[Pielorz 2008]{Pielorz08}
Pielorz, J., Ph.D. thesis, The University of Bonn,
{http://hss.ulb.uni-bonn.de/diss\_online/math\_nat\_fak/2008/pielorz\_jasmin/pielorz-abstract-engl.htm}

%
\bibitem[Riess et al. 1998]{Riess}
Riess, A.~G., et al., 1998, AJ, 116, 1009

%
\bibitem[Rimes \& Hamilton 2005]{RimesHamilton05}
Rimes, C.~D., Hamilton, A.~J.~S., 2005, MNRAS, 360, L82

%
\bibitem[Saito et al. 2008]{Saitoetal08}
Saito, S., Takada, M., Taruya, A., 2008, Phys.~Rev.~Lett., 100, 191301

%
\bibitem[Schneider06]{Schneider06}
Schneider, P., {\it Gravitational Lensing: Strong, Weak and Micro},
    Lecture Notes of 33rd Saas-Fee Advanced Course, G.~Meylan,
    P.~Jetzer \& P.~North (eds.), Springer-Verlag: Berlin (2006).

%
\bibitem[Schneider et al. 2002]{Schneideretal02}
Schneider, P., Van Waerbeke, L., Kilbinger, M., Mellier, Y., 
2002, A\&A, 396, 1

%
\bibitem[Scoccimarro et al 1999]{Scoccimarro99}
Scoccimarro, R., Zaldarriaga, M., Hui, L., 1999, ApJ, 527, 1

%
\bibitem[Scoccimarro et al. 2001]{Scoccimarro01}
Scoccimarro, R., Sheth, R., Hui, L., Jain, B., 2001, ApJ, 546, 20



%
\bibitem[Sefusatti et al. 2006]{Sefusatti06}
Sefusatti, E., Crocce, M., Pueblas, S., Scoccimarro, R., 2006, 
Phys.~Rev.~D, 74, 023522 

%
\bibitem[Seljak 2000]{Seljak00}
Seljak, U., 2000, MNRAS, 318, 203

%
\bibitem[Seljak \& Zaldarriaga 1996]{CMBFAST}
Seljak, U., Zaldarriaga, M., 1996, ApJ, 469, 437

%
\bibitem[Semboloni et al. 2007]{Sembolonietal07}
Semboloni, E., Van Waerbeke, L., Heymans, C., Hamana, T., Colombi, S.,
White, M., Mellier, Y.,  2007, MNRAS, 375, L6

%
\bibitem[Smith et al. 2003]{Smithetal03}
Smith, R.~E., et al., 2003, MNRAS, 341, 1311 

%
\bibitem[Song \& Knox 2004]{SongKnox04}
Song, Y.-S., Knox, L., 2004, Phys.~Rev.~D, 70, 3510

%
\bibitem[Takada 2006]{Takada06}
Takada, M., 2006, Phys.~Rev.~D, 74, 043505 

%
\bibitem[Takada \& Bridle 2007]{TakadaBridle07}
Takada, M., Bridle, S., 2007,  New~J.~Phys., 9, 446

%
\bibitem[Takada \& Jain 2002]{TJ02}
Takada, M., Jain, B., 2002, MNRAS, 337, 875

%
\bibitem[Takada \& Jain 2003a]{TJ03a}
Takada, M., Jain, B., 2003a, MNRAS, 340, 580

%
\bibitem[Takada \& Jain 2003b]{TJ03b}
Takada, M., Jain, B., 2003b, MNRAS, 346, 949

%
\bibitem[Takada \& Jain 2004]{TJ04}
Takada, M., Jain, B., 2004, MNRAS, 348, 897

\bibitem[Takada \& White 2004]{TakadaWhite04}
Takada, M., White, M., 2004, ApJ, 601, L1

%
\bibitem[Takada et al. 2006]{Takadaetal06}
Takada, M., Komatsu, E., Futamase, T., 2006, Phys.~Rev.~D, 73, 083520

\bibitem[Takahashi et al. 2008]{Takahashietal08}
Takahashi, R., Yoshida, N., Takada, M., et al., in preparation.

%
\bibitem[Tegmark et al. 1997]{Temgarketal97}
Tegmark, M., Taylor, A.~N., Heavens, A.~F., 1997, ApJ, 480, 22

%
\bibitem[Tegmark et al. 2002]{Temgarketal02}
Tegmark, M., Hamilton, A.~S.~J., Xu, Y., 2002, MNRAS, 335, 887

%
\bibitem[White \& Hu 2000]{WhiteHu00}
White, M., Hu, W., 2000, ApJ, 537, 1

\end{thebibliography}
\end{document}